\documentclass[showpacs,twocolumn,superscriptaddress]{revtex4-1}
\usepackage{bm,graphicx,subfigure}
\usepackage{color}
\usepackage{epstopdf}

\usepackage[colorlinks=true, linkcolor=blue, citecolor=blue, urlcolor=black]{hyperref}

\begin{document}
\title{Ground state properties of trapped boson system with finite-range Gaussian repulsion: Exact diagonalization study}
\author{Mohd. Imran}\email{alimran5ab@gmail.com}
\affiliation{Department of Physics, Jamia Millia Islamia (A Central University), New Delhi 110025, India} 
\affiliation{Department of Physics, Lingaya's Vidyapeeth, Old Faridabad, Haryana 121002, India}
\author{M. A. H. Ahsan}\email{mahsan@jmi.ac.in}
\affiliation{Department of Physics, Jamia Millia Islamia (A Central University), New Delhi 110025, India} 
\date{\today}

\begin{abstract}
We use exact diagonalization to study an interacting system of $N$ spinless bosons with finite-range Gaussian repulsion, confined in a quasi-two-dimensional harmonic trap with and without an introduced rotation. 
The diagonalization of the Hamiltonian matrix using Davidson algorithm in subspaces of quantized total angular momentum $L_{z}$ is carried out to obtain the $N$-body lowest eigenenergy and eigenstate. 
To bring out the effect of quantum (Bose) statistics and consequent phase stiffness (rigidity) of the variationally obtained many-body wavefunction on various physical quantities, our study spans from few-body ($N=2$) to many-body ($N=16$) systems.  
Further, to examine the finite-range effect of the repulsive Gaussian potential on many-body ground state properties of the Bose-condensate, we obtain the lowest eigenstate, the critical angular velocity of single vortex state and the quantum correlation (measured) in terms of von Neumann entanglement entropy and degree of condensation.
It is found that for small values of the range (measured by the parameter $\sigma$) of Gaussian potential, the ground state energy increases for few-boson ($2\le N\le 8$) systems but decreases for many-boson ($N>8$) systems.
On the other hand for relatively large values of the range of Gaussian potential, the ground state energy exhibits a monotonic decrease, regardless of the number of bosons $N$.
For a given $N$, there is found an optimal value of the range of Gaussian potential for which the first vortex (with $L_{z}=N$) nucleates at a lower value of the rotational angular velocity $\Omega_{\bf c1}$ compared to the zero-range ($\delta$-function) potential.
Further, we observe that the inter-particle interaction and the introduced rotation are competing effects with latter being dominant over the former.
With increase in the range of Gaussian potential, the value of von Neumann entropy decreases and the degree of condensation increases implying an enhanced quantum correlation and phase rigidity.
\end{abstract}

% 03.65.-w  Quantum mechanics
% 03.65.Aa 	Quantum systems with finite Hilbert space
% 03.75.Hh 	Static properties of condensates; thermodynamical, statistical, and structural properties
% 05.30.Jp 	Boson systems
% 34.20.-b 	Interatomic and intermolecular potentials and forces, potential energy surfaces for collisions
% 45.50.Jf 	Few- and many-body systems
% 67.85.-d 	Ultracold gases, trapped gases

\maketitle

\section{Introduction}
\label{intro}
The experimental realization of Bose-Einstein condensation (BEC) at nanokelvin temperatures in alkali Bose atomic vapors \cite{aem95,dma95,bst95} offers a possibility to examine the role of various energy scales (associated with size, zero-point motion, interaction, rotation etc.) in the physics of interacting bosons confined in a trap with an externally impressed rotation \cite{isw99,dgp99,fs01,ajl01}.
Contrary to existing bulk systems like liquid $^4$He where several of these energy scales overlap, the new mesoscopic systems that provide decisive control (in the laboratory) over their physical parameters \cite{ps02,jfa03} such as density, effective dimensionality and particle-particle interaction \cite{ias98,cgj10}, enables one to separate and study their effects on the quantum many-body states \cite{bdz08,coo08,fet09}.
Further, with recent advances in BEC on optical lattices \cite{mo06,bpt10,swe10} and microchip traps \cite{lsh03,fz07}, few-body systems have become an experimental reality and attracted particular attention of theorists \cite{yl07,vie08,srh10,blu12,imp15,msp17,kmk17}.
The similar Rydberg multi-wave mixing processes enhanced by the atomic coherence are well known theoretically and experimentally \cite{zgg18,zzy15}.
In atomic or atomic-like medium, the multi-mode correlation and squeezing of parametrically amplified multi-wave mixing have also been discussed in recent years \cite{zwn11,ljz17}.
\\
\indent
For atomic gas systems at low-enough temperatures with short-ranged atom-atom interaction, the interparticle interaction is usually described by zero-range ($\delta$-function) potential \cite{dgp99,fs01,ajl01,ber98,wg00,cwg01,lhd10}. 
This approach is useful in case of one dimensional systems as $\delta$-function potential is well understood and easy to handle in 1D.
Unfortunately, such an interaction potential in 2D is not self-adjoint \cite{ber98} and cannot be used in a manner analogous to its one dimensional counterpart.
Using configuration interaction, it is shown in Ref.~\cite{eg99} that a $\delta$-function potential is not suitable as a replacement for the two-body interaction in exact theories.
Indeed a regularized 2D contact potential is analytically tractable \cite{far10} but harder to handle numerically.
Further as the number of particles is increased beyond two, the complexity of the quantum states increases quickly and analytical treatment becomes intractable, leaving one only with a numerical recourse \cite{bp99,wg00}. 
It has been demonstrated that in order to tackle the limit of zero-range interaction numerically \cite{dka13}, a prohibitively large Hilbert space is required to obtain the many-body ground state. 
Thus to avoid regularization problem of the zero-range $\delta$-function potential in two dimensions \cite{eg99} for numerical many-body simulation, one usually prefers a smooth, short-range, model interaction potential.
This leads one to use the finite-range Gaussian type of potential as a model interparticle interaction in the trapped many-body quantum systems \cite{cfa09,dka13,kls14,bkc15,iasl15,flc15,sk16,iactp16,wtc17,iajpb17,bca18,jcb18}.
It is pertinent to mention here that in addition to being physically more realistic, the finite-range Gaussian type of potential~(\ref{gip}) is expandable within a finite subspace of single-particle basis functions \cite{hua87,motiv} and hence computationally more feasible \cite{cfa09,dka13}.
\\
\indent
A better control over interparticle interaction, characterized independently by the strength $\&$ range of interaction, and being expandable within finite number of basis functions of the Hilbert space \cite{cfa09,dka13}, are few of the advantages of the finite-range Gaussian shaped potential over the usual $\delta$-function potential. 
Besides, most of the mean field as well as many-body calculations on harmonically confined interacting bosons with an externally impressed rotation have used only the lowest-Landau-levels (LLL), in which bosons occupy single-particle states with radial quantum number $n_{r}=0$ and angular momentum quantum number $m$ taking only  positive sign. 
There has been several exact diagonalization studies \cite{ahs01,lhc01_pra} wherein it has been argued that for slow rotating~\cite{rot} and moderately to strongly interacting~\cite{int} bosons, it becomes necessary to consider the single-particle states beyond LLL approximation with $n_{r}=0,1,\dots$, while $m$ is allowed to take positive as well as negative values in constructing the many-body basis states. 
We are thus motivated to study the many-body dynamics beyond LLL approximation in moderately interacting system with finite-range Gaussian type of potential.
\\
\indent
In the present work, our aim is to investigate various quantum mechanical ground state properties of rotating system of $N$ $\left(2 \le N \le 16 \right)$ spinless bosons interacting via finite-range repulsive Gaussian type of potential (of large $s$-wave scattering length) in a quasi-two-dimensional (quasi-2D) harmonic trap. 
The exact diagonalization of the many-body Hamiltonian matrix is carried out in given subspaces of quantized total angular momentum $L_{z}$ using Davidson iterative algorithm \cite{dav75} to obtain the lowest-energy eigenstates in the co-rotating frame.
We focus on angular momentum regime $0 \le L_{z} \le 2N $ and are mainly interested in exploring the system in vicinity of single vortex ($L_{z}=N$) state. 
We examine the role of the Gaussian range of interaction on various quantities of interest that are experimentally accessible such as the ground state energy,  von Neumann entanglement entropy, degree of condensation, fragmentation if any and critical angular velocity $\Omega_{\bf c1}$ of the first vortex state $L_{z}=N$.
The results obtained demonstrate that the use of repulsive interparticle Gaussian shaped potential has significant effect on the many-body ground state properties. 
\\
\indent 
This paper is organized as follows.
In Sec.~\ref{model}, we describe the model Hamiltonian of a rotating Bose system with finite-range Gaussian shaped interaction in quasi-2D harmonic trap.
We then introduce the single-particle reduced density matrix and the criterion for the existence of BEC, von Neumann entanglement entropy and degree of condensation. 
In Sec.~\ref{results}, we present our exact diagonalization results on a system of $N$ bosons, to examine the finite-range effect of interaction on many-body ground state properties. 
We first examine the non-rotating system with $L_{z}=0$ followed by the rotating system in different subspaces of $L_{z}>0$ focusing on the first vortex state with $L_{z}=N$.
Finally in Sec.~\ref{conc}, we summarize the main results of the present study.

\section{Theoretical Model}
\label{model}
\subsection{The System and The Hamiltonian}
We consider a system of $N$ interacting spinless bosons each of mass $M$, trapped in a harmonic potential
$V({\bf r})={\frac{1}{2}}M\left(\omega_{\perp}^{2} {r}^{2}_{\perp} + \omega_{z}^{2} {z}^{2}\right)$.
The trap with $x$-$y$ symmetry is taken to rotate about the $z$-axis with angular velocity ${\bm{\widetilde{\Omega}}}\equiv \widetilde{\Omega} \hat{e}_{z}$.
Here ${r}_{\perp}=\sqrt{x^{2} +y^{2}}$ is the radial distance of the particle from the trap center; $\omega_{\perp}$ and $\omega_{z}$ are the radial and the axial frequencies respectively, of the harmonic confinement. 
The axially symmetric trapping potential $V\left({\bf r}\right)$ is assumed to be highly anisotropic with $\lambda_{z}\equiv \omega_{z} / \omega_{\perp}\gg 1$.
The confined system is thus effectively quasi-two-dimensional. We chose $\hbar \omega_{\perp}$ as the unit of energy and $a_{\perp} = \sqrt{\hbar/{M \omega_{\perp}}}$ the corresponding unit of length. 
Introducing $\Omega \equiv \widetilde{\Omega}/{\omega_{\perp}}$ $(\leq 1)$ as the dimensionless angular velocity and $L_{z}$ (scaled by $\hbar$) the $z$ projection of the total angular momentum operator, the many-body Hamiltonian in the co-rotating frame is given as ${H}^{rot}={H}^{lab}-\Omega L_{z}$ where
\begin{equation}
H^{lab} = \sum_{j=1}^{N} \left[-\frac{1}{2} \bm{\nabla}^{2}_{j} + \frac{1}{2} {\bf r}_{j}^{2} \right] + \frac{1}{2} \sum_{i\neq j}^{N} U \left({\bf r}_{i},{\bf r}_{j}\right)
\label{mbh}
\end{equation} 
The first two terms in the Hamiltonian~(\ref{mbh}) correspond to the kinetic energy and confining potential respectively.
The third term $U\left({\bf r}_{i},{\bf r}_{j}\right)$, arises from the inter-particle interaction described by the Gaussian shaped potential \cite{cfa09,dka13,kls14,bkc15,iasl15,flc15,sk16,iactp16,wtc17,iajpb17,bca18,jcb18}
\begin{equation}
U \left({\bf r}_{i},{\bf r}_{j}\right) = \frac{\mbox{g}_{2}} {2\pi{\sigma^{2}}}
\exp{\left[ -\frac{\left({\bf r}_{i}-{\bf r}_{j}\right)^{2}}{2\sigma^{2}} \right]}
\label{gip}
\end{equation}
with parameter $\sigma$ (scaled by $a_{\perp}$) being the effective range of the potential and the dimensionless parameter $\mbox{g}_{2}\equiv 4\pi {a_{s}}/{a_{\perp}}$ being a measure of the strength of interaction where $a_{s}$ is the $s$-wave scattering length taken to be positive $\left(a_{s}>0\right)$ so that the effective interaction is repulsive. 
In the limit $ \sigma \rightarrow 0$, the Gaussian type of potential in Eq.~(\ref{gip}) reduces to the zero-range $\delta$-function potential $\mbox{g}_{2}\delta\left({\bf r}-{\bf r}^{\prime}\right)$ used in earlier studies \cite{dgp99}. In the opposite limit of $\sigma \rightarrow \infty$, the Gaussian  potential $U \left({\bf r}_{i},{\bf r}_{j}\right)\rightarrow 0$ for a given inter-particle separation $\left|{\bf r}_{i}-{\bf r}_{j}\right|$ and a finite strength of interaction $\mbox{g}_{2}$.
The value of interaction range $\sigma$ is varied from zero to the system size $\sim a_{\perp}$.
For a given value of $\sigma$, the $s$-wave scattering length in the parameter $\mbox{g}_{2}=4\pi {a_{s}}/{a_{\perp}}$ is adjusted in such a way that $Na_{s}/a_{\perp}$ becomes relevant to experimental values \cite{dgp99}.
\\
\indent
Since the system being studied here is rotationally invariant in the $x$-$y$ plane, the $z$-component of the total angular momentum is a good quantum number leading to block diagonalization of the Hamiltonian matrix in subspaces of $L_{z}$ \cite{ahs01}. 
To obtain the many-body eigenstates, we carry out diagonalization of the Hamiltonian matrix in different subspaces of $L_{z}$ with inclusion of lowest as well as higher Landau levels of the single-particle basis states in constructing the $N$-body basis states \cite{ahs01,lhc01_pra,iajpb17}.
\subsection{The single-particle basis states}
The single-particle basis functions are chosen to be the eigenfunctions of the non-interacting single-particle Hamiltonian
\begin{equation}
H_{sp} = \frac{1}{2} \left(-{\bf\nabla}^{2}_{\perp} + r_{\perp}^{2}\right) - {\Omega} \, {\ell_{z}} + \frac{1}{2} \left(-{\bf\nabla}^{2}_{z} + \lambda_{z}^{2} z^{2}\right),
\label{sph1}
\end{equation}
identified as the harmonic oscillator Hamiltonian with an externally impressed rotation about the $z$-axis given   the single-particle angular momentum $\ell_{z}$. The eigensolutions of ${H}_{sp} \, {u}_{n,m,n_{z}}\left({\bf r}\right)={\epsilon}_{n,m,n_{z}}{u}_{n,m,n_{z}} \left({\bf r}\right)$, in dimensionless form, are known to be:
\begin{eqnarray} 
&&{\epsilon}_{n,m,n_{z}} = \left(n+1-m{\Omega}\right) + {\lambda}_{z}\left(n_{z}+{1}/{2}\right)\nonumber \\
&&{u}_{n,m,n_{z}} \left({\bf r}\right) =
\sqrt{\frac{\left(\frac{1}{2}\left\{ n-|m| \right\} \right)! \, \sqrt{{\lambda}_{z}/\pi^{3}}}{\left(\frac{1}{2}\left\{ n+|m| \right\} \right)! \ 2^{n_{z}} \ n_{z}!}}\ e^{-\left({r_{\perp}^{2}+ {\lambda}_{z} z^{2}}\right)/2} \nonumber \\
&&~~~~~~~~~~~~~~~~ \times e^{im\phi } \ {r^{|m|}_{\perp }}\, 
L^{|m|}_{\frac{1}{2}(n-|m|)} \left(r_{\perp }^{2}\right) H_{n_{z}}\left({\lambda}_{z}z^{2}\right)
\label{sps1}
\end{eqnarray}
where $n=2n_{r}+|m|$ with the radial quantum number $n_{r}$ and the single-particle angular momentum quantum number $m$. Here $L^{|m|}_{\frac{1}{2}(n-|m|)}\left(r_{\perp }^{2}\right)$ is the associated Laguerre polynomial and $H_{n_{z}}\left({\lambda}_{z}z^{2}\right)$ the Hermite polynomial.  
With $\lambda_{z}\gg 1$, the system is taken to be quasi-2D and hence there is practically no excitation along the $z$-axis. We, therefore, set $n_{z}=0$ in Eq.~(\ref{sps1}) implying that all the particles occupy the lowest-energy state ${u}_{0}(z)=({\lambda}_{z}/\pi)^{1/4}\ e^{-{{\lambda}_{z} z^{2}}/2}$ of $z$ co-ordinate degree of freedom, thereby reducing
 Eq.~(\ref{sps1}) to 
\begin{eqnarray} 
&&\epsilon_{n,m}=\left(n+1-m {\Omega}\right)+ {\lambda_{z}}/{2}~,~~\mbox{with}~n=2n_{r}+|m| \nonumber \\
&&{u}_{n,m} \left({\bf r}\right)=
\sqrt{\frac{\left(\frac{1}{2}\left\{ n-|m| \right\} \right)!}{\left(\frac{1}{2}\left\{ n+|m| \right\} \right)!}\sqrt{\frac{{\lambda}_{z}}{\pi^{3}}}}\ e^{-\left({r_{\perp}^{2}+ {\lambda}_{z} z^{2}}\right)/2}
\nonumber \\
&&~~~~~~~~~~~~~~ \times e^{im\phi }\ {r_{\perp }^{|m|}}\ 
L^{|m|}_{\frac{1}{2}(n-|m|)} \left(r_{\perp }^{2}\right).
\label{sps2}
\end{eqnarray}
Restricting to $n_{r}=0$ and taking $m \geq 0$ corresponds to the lowest-Landau level (LLL) approximation. 
Taking $n_{r}=0,1,2,\dots $ and allowing $m$ to take positive as well as negative values corresponds to going beyond LLLs \cite{ahs01,lhc01_pra}. 
\\
\indent
In the present work, we employ beyond LLL approximation by including the lowest as well as higher Landau levels of the single-particle basis states ${u}_{n,m}\left({\bf r}\right)$ with $n_{r}=\frac{1}{2} \left( n-|m|\right)= 0, 1 $ and $m=0,\pm 1,\pm 2,\pm 3,\cdots$.
For a system of $N$ Bose atoms in a given subspace of total angular momentum $L_{z}$, the single-particle angular momentum $m$ for the basis  functions $ u_{n,m}\left(r_{\perp },\phi\right)$ spanning the 2D $xy$-plane is chosen to be: $m=\ell_{z}-n_{b},\ \ell_{z}-n_{b}+1,\ \cdots \ell_{z}+n_{b}-1,\ \ell_{z}+n_{b}$.
It is convenient to define $\ell_{z} \equiv\left[L_{z}/N\right]$ where for real $x$ the symbol $\left[x\right]$ denotes the greatest integer less than or equal to $x$.
Here $n_{b}$ is some positive integer that one may chose to be 3, 4 or more depending on the strength of the interaction \cite{int} and the computational resources available ($n_{b}$ is a kind of the size of the single-particle basis chosen for calculation for a given value of $L_{z}$). 
Thus for instance, for $N=16$ and for the chosen subspace $L_{z}=37$, we get $\ell_{z} \equiv\left[{L}_{z}/{N}\right]=2$, and the single-particle angular momentum quantum number takes values $m=-1,0,+1,+2,+3,+4,+5$. Then, with $n_{r}=0,1$, the single-particle basis set turns out to be 
\begin{eqnarray*}
&&\left( u_{0,0},\ u_{1,+1},\ u_{2,+2},\ u_{3,+3},\ u_{4,+4},\ u_{5,+5},\right.\\ 
&& \left. u_{1,-1},\ u_{2,0},\ u_{3,+1},\ u_{4,+2},\ u_{5,+3},\ u_{3,-1}\right).
\end{eqnarray*}
The single-particle basis functions thus chosen are used to construct the many-particle basis states $\left\{\Phi_{\nu}\right\}$ in the variational trial function 
$\Psi = \sum_{\bm{\nu}}C_{\bm{\nu}}~\Phi_{\bm{\nu}}$ 
of the system for given value of total angular momentum $L_{z}$.
\subsection{Construction of many-body basis states}
The $N$-body basis states $\left\{\Phi_{\bm{\nu}} \left({\bf r}_{1},{\bf r}_{2},\dots , {\bf r}_{N}\right) \right\}$ in a given subspace of $L_{z}$ are constructed as the symmetrized products of a finite number of single-particle basis functions ${u}_{n,m}\left({\bf r}\right)$ defined in Eq.~(\ref{sps2}).
The many-body index $\bm{\nu}\equiv \left(\nu_{\bf 0},\nu_{\bf 1},\dots,\nu_{\bf k},\dots, \nu_{\bf j}\right)$ labeling the many-body basis function $\Phi_{\bm{\nu}} \left({\bf r}_{1},{\bf r}_{2},\dots, {\bf r}_{N} \right)$ stands for a set of single-particle quantum numbers $\left\{ {\bf k} \equiv \left(n,m\right)\right\}$ and their respective occupancies $\left\{ \nu_{\bf k}\right\}$.
In the second-quantized notation, the Bose field operator is expanded in terms of single-particle basis functions  as $\hat{\psi} \left({\bf r}\right) = \sum_{\bf k}\hat{b}_{\bf k} u_{\bf k} \left({\bf r}\right)$ where $\hat{b}_{\bf k} (\hat{b}^{\dagger}_{\bf k})$ is the annihilation(creation) operator for the Bose quanta in state $\bf k$, obeying commutation rules.
In occupation-number representation, the $N$-body basis function $\vert \Phi_{\bm{\nu}} \rangle$ in a given subspace of $L_{z}$ is written  in second-quantized form as:  
\begin{equation}
\vert \Phi_{\bm{\nu}} \rangle \equiv  \prod_{\bf k=0}^{\bf j}
\frac{1}{\sqrt{\nu_{\bf k}!}}{\left(\hat{b}^{\dagger}_{\bf k}\right)}^{\nu_{\bf k}} \vert \mathrm{vac} \rangle 
\equiv  \vert \nu_{\bf 0}\ \nu_{\bf 1}\cdots \nu_{\bf k} \cdots \nu_{\bf j}\rangle 
\label{nbb}
\end{equation}
with
$\sum_{{\bf k}={\bf 0}}^{\bf j}\nu_{\bf k}={{N}}$ and
$\sum_{{\bf k}={\bf 0}}^{\bf j}m_{\bf k}\nu_{\bf k}={L_{z}}$ where ${\bf k}=(n_{\bf k},m_{\bf k})$.
With these constraints on $\{\nu_{\bf k} , m_{\bf k}\}$ , only the most relevant Fock states (spanning the active Fock space), constructed from the full single-particle basis set, are included.
This procedure reduces the dimension of the Hamiltonian matrix to the order of ${10}^{2}-{10}^{5}$, amenable to diagonalization on a small workstation of $3.20$ GHz processor.
\subsection{Determination of variational many-body wavefunction}
In Rayleigh-Ritz scheme \cite{gok88} employed here, the $N$-body variational wavefunction $\Psi({\bf r}_{1},{\bf r}_{2},\dots,{\bf r}_{N})$ for a given $N$ and $L_{z}$, is constructed as 
\begin{equation}
\Psi\left({\bf r}_{1},{\bf r}_{2},\dots,{\bf r}_{{N}}\right) = \sum_{\bm{\nu}} {{C}_{\bm{\nu}}} \, \Phi_{\bm{\nu}} \left({\bf r}_{1},{\bf r}_{2},\dots,{\bf r}_{N} \right)
\label{nbf}
\end{equation}
where $\left\{ C_{\bm{\nu}} \right\}$ are the variational parameters to be determined.
Having constructed the active Fock states as outlined in the previous subsection, one may calculate the matrix elements and set up the eigenvalue problem $HC=EC$ where $H$ is the Hamiltonian matrix and $C$ is the column matrix formed by $\left\{ C_{\bm{\nu}} \right\}$. 
The eigenvalue problem is solved iteratively using Davidson algorithm \cite{dav75} to determine the many-body ground and low-lying excited eigenstates. 
\\
\indent
For a system of $N$ bosons confined in a harmonic trap rotating with angular velocity $\Omega$, the thermodynamic  equilibrium corresponds to minimizing the Helmholtz free energy $F(T,V,N) \equiv E(S,V,N)-TS$ given by $e^{-\beta F}=\mbox{Tr}\left[ e^{-\beta \left(H^{lab}-\Omega L_{z}\right)}\right]$.
At zero temperature, it reduces to $E =\left\langle \Psi_{0} \left|\left({H^{lab}-\Omega L_{z}}\right)\right|\Psi_{0}\right\rangle$, where $\Psi_{0}$ is the variationally obtained ground state with total angular momentum $L_{z}$, in the non-rotating (laboratory) frame.
Therefore, the many-body Hamiltonian ${H}^{lab}$ in Eq.~(\ref{mbh}) is diagonalized in given subspaces of $L_{z}$ to obtain the energy $E^{rot}\left(L_{z},\Omega\right)=E^{lab}\left(L_{z}\right)-\Omega L_{z}$ in the co-rotating frame.  
This is equivalent to minimizing $E^{lab}\left( L_{z}\right)$ subject to the constraint that the system has angular momentum expectation value $L_{z}$ with angular velocity $\Omega$ identified as the corresponding Lagrange multiplier.

\subsection{Physical Quantities of Interest}
\paragraph*{Single-particle reduced density matrix (SPRDM).}
Starting from the normalized $N$-body ground state wavefunction $\Psi_{0}({\bf r}_{1},{\bf r}_{2},{\bf r}_{3},\dots,{\bf r}_{N})$, obtained variationally through exact diagonalization; 
one can determines the  single-particle reduced density matrix $\rho({\bf r},{\bf r}^{\prime})$, by integrating out the degrees of freedom of $\left(N-1\right)$ particles. 
Thus
\begin{eqnarray}
\rho \left({\bf r},{\bf r}^{\prime}\right) &=&\int \int \dots \int d{\bf r}_{2}\ d{\bf r}_{3}\dots d{\bf r}_{{N}} \nonumber \\
&&\times\ \Psi_{0}^{\ast}({\bf r},{\bf r}_{2},{\bf r}_{3} \dots,{\bf r}_{{N}})\ \Psi_{0}({\bf r}^{\prime},{\bf r}_{2},{\bf r}_{3},\dots,{\bf r}_{{N}}) \nonumber \\
&\equiv & \sum_{{\bf n},{\bf n}^{\prime}}\ \rho_{_{{\bf n},{\bf n}^{\prime}}}\ u^{\ast}_{\bf n}\left({\bf r}\right)u_{{\bf n}^{\prime }}\left({\bf r}^{\prime }\right).  
\end{eqnarray}
The above formulation is expressed in terms of single-particle basis functions $u_{\bf n}\left({\bf r}\right)$ with quantum number ${\bf n}\equiv \left(n,m\right)$.
Being hermitian, it can be diagonalized to give 
\begin{equation}
\rho \left({\bf r},{\bf r}^{\prime}\right) = \sum_{\mu }\lambda_{\mu } \ \chi^{\ast}_{\mu }\left({\bf r}\right)
\chi_{\mu } \left({\bf r}^{\prime }\right),
\label{spd}
\end{equation} 
with normalization $\sum_{\mu }\lambda_{\mu }=1$ and $ \chi_{\mu }\left({\bf r}\right) \equiv \sum_{\bf n} c^{\mu }_{\bf n} \, {u}_{\bf n}\left({\bf r}\right)$ where each $\mu$ defines a fraction of the BEC.
The $\left\{\lambda_{\mu }\right\}$ are the eigenvalues, ordered as $ 1\geq\lambda_{1}\geq\lambda_{2}\geq\cdots\lambda_{\mu }\cdots \geq 0 $, and $\left\{ \chi_{\mu }\left({\bf r}\right)\right\}$ are the corresponding eigenvectors of the single-particle reduced density matrix $\rho\left({\bf r},{\bf r}^{\prime }\right)$.
\paragraph*{Degree of condensation.}
It is to be noted that the usual definition of condensation for a macroscopic system, given by the largest eigenvalue $\lambda_{1}$ of the single-particle reduced density matrix Eq.~(\ref{spd}), is not appropriate for systems with small number of particles being studied here \cite{dbo07,iasl15}.
For example, in the absence of condensation, there is no macroscopic occupation of a single quantum state and all eigenvalues of the single-particle reduced density matrix are of the same order, such a definition would imply a condensate though with small magnitude. 
To circumvent this situation, one introduces a quantity which is sensitive to the loss of macroscopic occupation called the degree of condensation \cite{dbo07} defined as
\begin{equation}
C_{d} =\lambda_{1}-\bar{\lambda}
\label{doc}
\end{equation}
where $\bar{\lambda}=\frac{1}{p-1}\sum_{\mu=2}^{p} \lambda_{\mu}$ is the mean of the rest of eigenvalues. 
It can be seen that the degree of condensation, defined as in Eq.~(\ref{doc}), approaches zero for equal eigenvalues, as one would expect.
\paragraph*{Von Neumann entanglement entropy.}
In order to measure the quantum correlation of a many-body ground state, we calculate von Neumann entanglement entropy \cite{py01,lgc09,lf10} defined as 
\begin{equation}
S_{1}=-\mbox{Tr}\left(\hat{\rho} \, \ln \hat{\rho} \ \right)
\end{equation}
with $\hat{\rho}$ being the single-particle reduced density operator obtained from the many-body ground state wavefunction.
In terms of eigenvalues $\left\{\lambda_{\mu }\right\}$ and eigenfunctions $\left\{\chi_{\mu }\left({\bf r}\right)\right\}$, then, the von Neumann entropy is evaluated explicitly as
\begin{equation}
S_{1}=-\sum_{\mu}\lambda_{\mu} \ln \lambda_{\mu}
\label{qent}
\end{equation}
in subspaces of total angular momentum $L_{z}$.
The entropy $S_{1}$ also provides information about the degree of condensation $C_{d}$.
For instance, for a perfect BEC, the value of $C_{d}$ approaches to unity  correspondingly the value of $S_{1}$ is zero, from Eq.~(\ref{doc}) and (\ref{qent}) respectively, since all bosons assume to occupy one and the same mode that is macroscopic eigenvalue $\lambda_{1} \sim 1$ and $\lambda_{\mu} \sim 0$ for ${\mu}>2$. 
As the condensate depletes (due to rotation or interaction), with more than one eigenvalue $\left\{\lambda_{\mu} \right\}$ becoming non-zero, $S_{1}$ increases and correspondingly $C_{d}$ decreases.
\section{Numerical Results and Discussion}
\label{results}
In the discussion below a system with $2 \le N < 8$ bosons where the effect of quantum statistics is insignificant is referred to as few-boson system whereas a system with $ N \ge 8 $ bosons where the effect of quantum statistics begins to show up is referred to as many-boson system.\\
\indent
To be concrete, we consider an interacting system of $2 \le N \le 16$ Bose atoms of $^{87}$Rb confined in an axially symmetric harmonic trap.
The radial trapping frequency is taken to be $\omega_{\perp}=2\pi \times 220 \, $Hz corresponding to the trap length $a_{\perp}=\sqrt{{\hbar}/{M\omega_{\perp}}}=0.727 \, \mu$m and the  aspect ratio of the trap $\lambda_{z} \equiv \omega_{z}/ {\omega_{\perp}}=\sqrt{8}$, so that the system has small extension $a_{z}=\sqrt{\hbar/M\omega_{z}}=a_{\perp}\lambda_{z}^{-{1}/{2}}$ in the $z$-direction, resulting in a quasi-2D system.
Recent advancements in atomic physics have made it possible to tune the strength of interaction (from weak to strong) in ultracold atomic vapors using Feshbach resonance \cite{ias98,cgj10}. 
For the many-body system being studied here, the characteristic energy scale for the interaction is determined by the dimensionless parameter $\left(Na_{s}/a_{\perp}\right)$ \cite{dgp99}.
Owing to increasing dimensionality of the Hilbert space with number of particles $N$, making computation impractical for $N$ more than a few hundred particles. 
We in our calculation here vary the s-wave scattering length $a_{s}$ to achieve suitable value of $\left(N a_{s}/a_{\perp}\right)$ relevant to experimental situation. 
In the results presented here, the parameters of repulsive interaction in Eq.~(\ref{gip}) have been chosen as $0 \leq \sigma \leq 0.9$ (in units of $a_{\perp}$) and $a_{s}=10000 \, a_{0}$ (where $a_{0}=0.05292 \, $nm is the Bohr radius) so that the parameter $(Na_{s}/a_{\perp}) \sim 1$ corresponds to moderate interaction \cite{ahs01,lhc01_pra}.
With this choice, the value of the dimensionless parameter $\mbox{g}_{2}={4\pi a_{s}}/{a_{\perp}}$ (measure of interaction strength) turns out to be $9.151$.
From now on, we fix the value of interaction strength $\mbox{g}_{2}=9.151$ and vary the Gaussian range $\sigma$ from zero (corresponding to $\delta$-interaction limit) to a value determined by the system size $\sim a_{\perp}$.
The $N$-body variational ground state wavefunction beyond lowest-Landau-level (LLL) approximation \cite{ahs01,lhc01_pra} is obtained through exact diagonalization of the Hamiltonian matrix for each of the subspaces of quantized total angular momentum $0\leq {L}_{z} \le 2N$ using Davidson iterative algorithm \cite{dav75}.
The results obtained with repulsive Gaussian shaped interaction~(\ref{gip}) allow us to study the effect of the range $\sigma$ of interparticle interaction on various quantum mechanically interesting properties of the Bose-condensed gas discussed in the following.

\subsection{Non-rotating System} 
\begin{figure}[!htb]
{\includegraphics[width=0.98\linewidth]{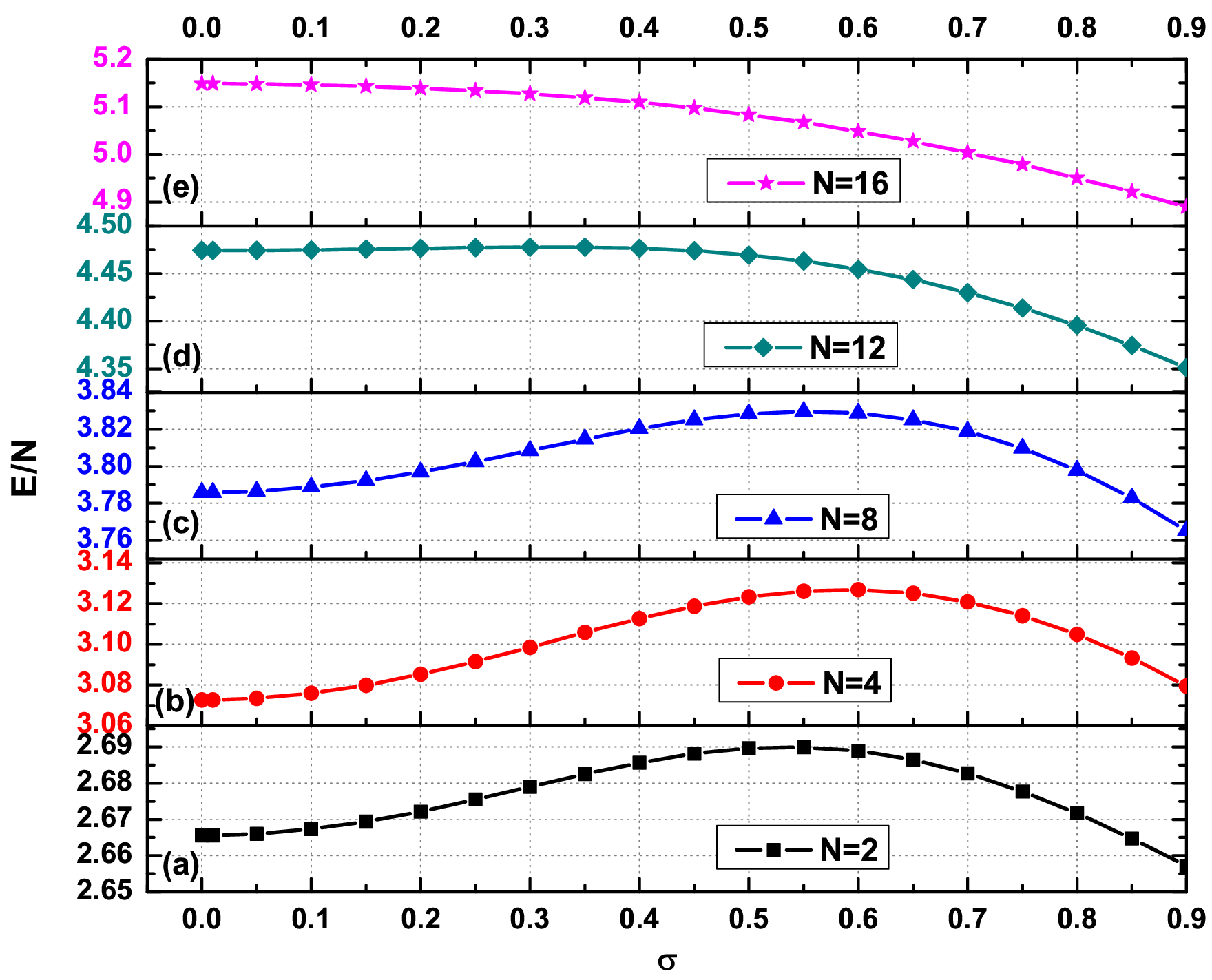}}
\caption{\label{fig:l0_ens}(Color online) The non-rotating $L_{z}=0$ state: ground state energy per particle $E/N$ (in units of $\hbar \omega_{\perp}$) {\it versus} the interaction range $\sigma$ (in units of $a_{\perp}$) of the Gaussian potential~(\ref{gip}) with $\mbox{g}_{2}=9.151$ for a system of (a) $N=2$ (b) $N=4$ (c) $N=8$ (d) $N=12$ and (e) $N=16$ bosons. The limiting case $\sigma \rightarrow 0$ corresponds to contact ($\delta$-function) potential.}
\end{figure}

When the system is subjected to external rotation with angular velocity $0\le\Omega <\Omega_{c1}$ (where $\Omega_{c1}$ is the critical angular velocity of the first vortex state), the total angular momentum state $L_{z}=0$ corresponds to the many-body ground state (with lowest energy in the co-rotating frame). 
Here, we examine the non-rotating $N$-boson $\left(2\le N \le 16\right)$ ground state in $L_{z}=0$ subspace as the range of interaction is varied over $0 \le \sigma \le 0.9$ for the repulsive Gaussian  potential~(\ref{gip}).
\paragraph*{Many-body ground state energy.}
We first explore the dependence of N-body ground state energy on the parameter $\sigma$.
In Fig.~\ref{fig:l0_ens}, we present the variation of ground state energy per particle $\left(E/N\right)$ with the interaction range $\sigma$ for $N=2,\,4,\,8,\,12$ and $16$ bosons in $L_{z}=0$ subspace. 
For $N=2$ bosons, we observe in Fig.~\ref{fig:l0_ens}(a) that the ground state energy initially increases for small values of $\sigma$ attaining a peak at $\sigma=0.55$ and then decreases monotonically as $\sigma$ is increased further.
When the interaction range $\sigma $ is much greater than the system size $a_{\perp}$ (not shown in the figure), the ground state energy is found to approach the non-interacting value {\it i.e.} the interaction energy becomes negligible.  
Our results on $\sigma$-dependence of the ground state energy of $N=2$ particle in small $\sigma$ limit is in good agreement with a recent work \cite{dka13}.
\\
\indent
To examine the effect of quantum statistics, we study $N$ bosons with repulsive Gaussian potential~(\ref{gip}) and find that there is a change in $\sigma$-dependence of the ground state energy as the number of bosons is increased.
For few boson systems with $N \le 4$, the effect of quantum statistics is insignificant and the results are qualitatively similar to those presented for $N=2$ bosons.
It is seen from Figs.~\ref{fig:l0_ens}(a) and \ref{fig:l0_ens}(b) that the peak of the ground state energy per particle {\it versus} $\sigma$ plot shifts to larger values of $\sigma$ as the number of bosons is increased from $N=2$ to $N=4$.
For $N=8$, when the effect of quantum statistics begins to show up, the peak of the $E/N$ {\it versus} $\sigma$ plot shifts to smaller values of $\sigma$.
This reversal in trend of the peak position of $E/N$ {\it versus} $\sigma$ plot is due to quantum statistics as $N$ is increased from $2$ to $8$.
For $N>8$, the effect of quantum statistics becomes fully developed and the $E/N$ {\it versus} $\sigma$ plot exhibits a monotonic trend.
Accordingly, we observe from Fig.~\ref{fig:l0_ens}(d) for $N=12$ and Fig.~\ref{fig:l0_ens}(e) for $N=16$ that the ground state energy changes very little for small values of $\sigma$ but as $\sigma$ is increased further over the range $0.1 < \sigma \le 0.9$, the $E/N$ {\it versus} $\sigma$ plot exhibits a monotonic decrease showing no peak.  
The dependence of ground state energy on the interaction range $\sigma$ is quite similar (always decreasing with increasing $\sigma$) for relatively large values of $\sigma \ (< 1)$, regardless of the number of particles $N$.
It, however, shows departure for small values of $\sigma$, namely, the ground state energy increases for small number of particles $2 \le N \le 8$ but remains constant for $N>8$ as $\sigma$ is increased.
\paragraph*{Quantum correlation.}
To further analyze the effect of interaction range $\sigma$ of the Gaussian potential~(\ref{gip}) on the quantum mechanical correlation or phase coherence of the Bose-condensed gas, we calculate the single-particle reduced density matrix defined in Eq.~(\ref{spd}).
The many-body quantum correlation is then measured, here, in terms of the von Neumann entanglement entropy $S_{1}$ and the degree of condensation $C_{d}$ defined in Eq.~(\ref{qent}) and Eq.~(\ref{doc}), respectively.
\begin{figure}[!htb]
\subfigure[~$S_{1}$ {\it versus} $\sigma$]{\includegraphics[width=0.9\linewidth]{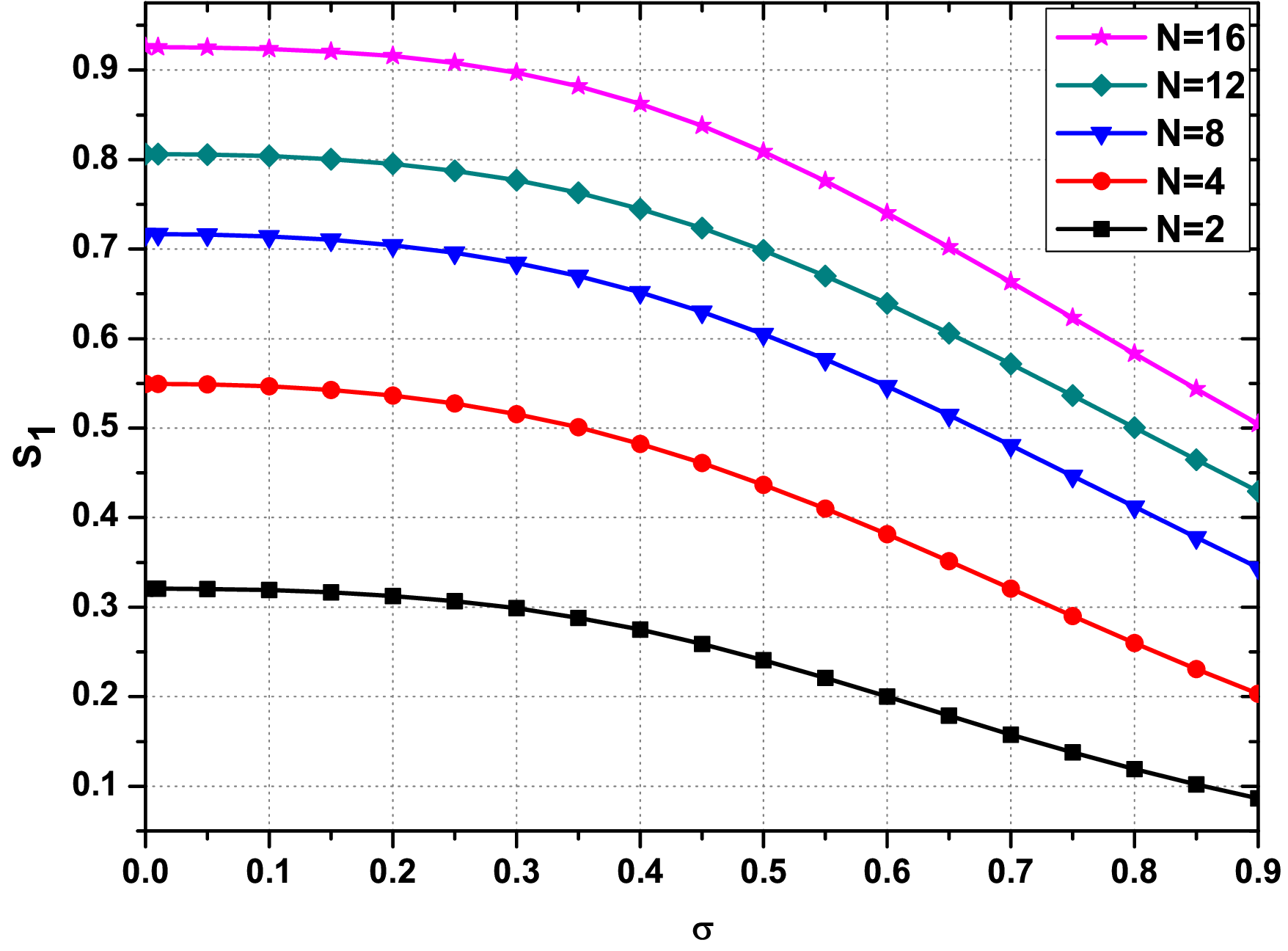}\label{fig:l0_qent}}
\subfigure[~$C_{d}$ {\it versus} $\sigma$]{\includegraphics[width=0.9\linewidth]{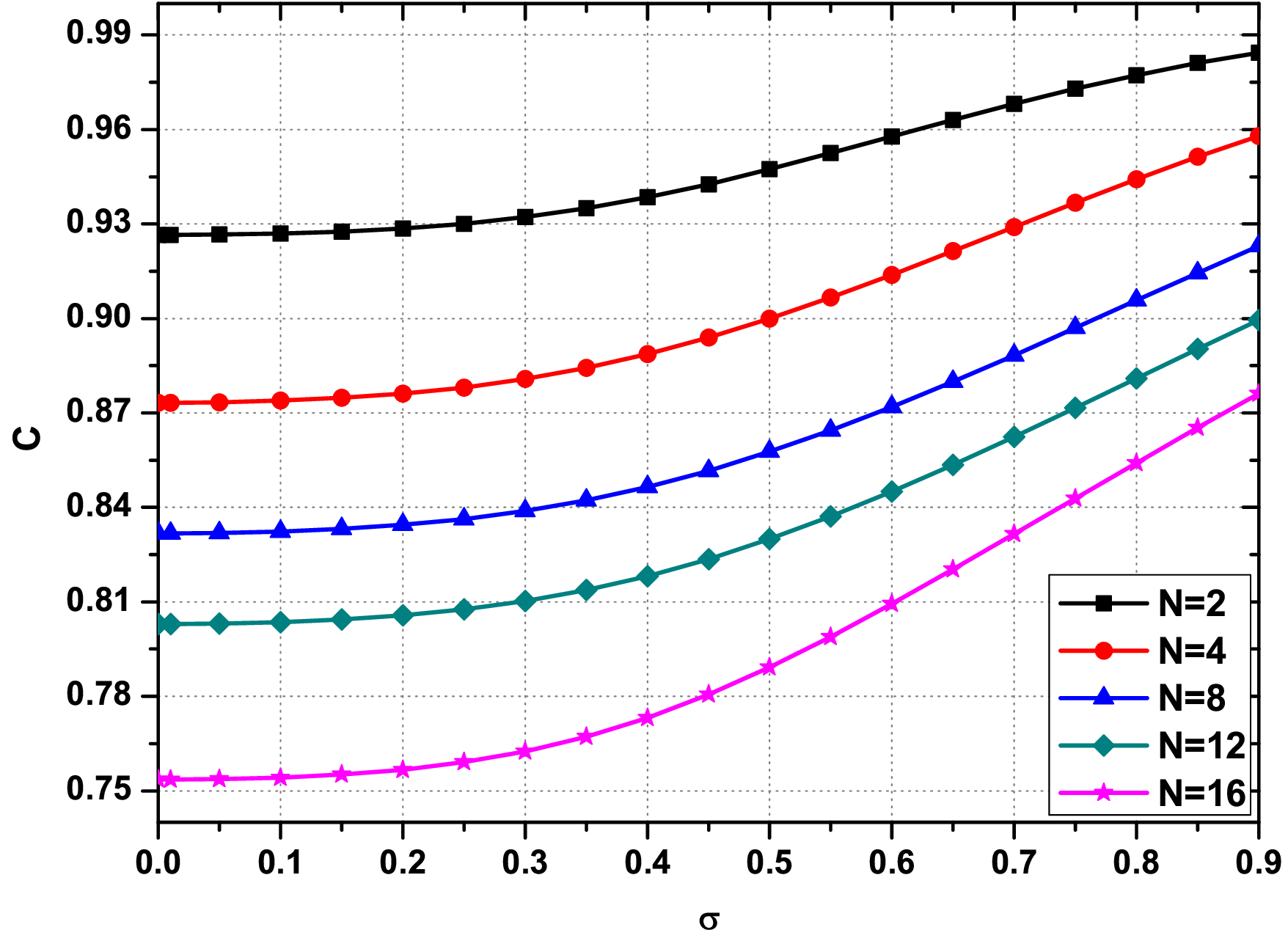}\label{fig:l0_doc}}
\caption{\label{fig:qentdoc}(Color online) For the non-rotating system corresponding to $L_{z}=0$ subspace, the $N$-body ground state of $N=2$~(black squares), $4$~(red circles), $8$~(blue triangles), $12$~(green rhombus) and $16$~(pink stars) bosons interacting via Gaussian potential~(\ref{gip}) with $\mbox{g}_{2}=9.151$; (a) von Neumann entropy $S_{1}$ and (b) degree of condensation $C_{d}$ as a function of the interaction range $\sigma$ (in units of $a_{\perp}$).}
\end{figure}
\\
\indent
We first examine the dependence of von Neumann entropy $S_{1}$ on the interaction range $\sigma$ of the Gaussian potential~(\ref{gip}).
The results are presented in Fig.~\ref{fig:l0_qent} where the values of entropy $S_{1}$ are plotted as a function of $\sigma$ for a system of $N=2,\,4,\,8,\,12$ and $16$ bosons in the angular momentum subspace $L_{z}=0$.
For a given $N$, the value of $S_{1}$ decreases with increase in $\sigma$ and becomes linear at large values of $\sigma \ (< 1)$.
The dependence of $S_{1}$ on $\sigma$ is qualitatively similar for all $N$-boson systems studied here.
It is interesting to note that for relatively large particle numbers $N \ge 8$, the entropy $S_{1}$ decreases sharply with increasing $\sigma$ as the quantum correlation increases with increasing $N$.    
Accordingly, $S_{1}$ {\it versus} $\sigma$ plot for $N=16$, the largest system considered in our case, is the steepest.
\\
\indent
To enlighten the discussion, the above observation of many-body quantum correlation is further supported by the results on the degree of condensation.
In Fig.~\ref{fig:l0_doc}, we present degree of condensation $C_{d}$ {\it versus} $\sigma $ for a system of $N=2,\,4,\,8,\,12$ and $16$ bosons.
For a given $N$, it is seen from the figure, the value of $C_{d}$ increases with increase in $\sigma$ and the variation becomes linear for larger values of $\sigma \ (< 1)$.
Consequently, with increasing $\sigma$, the degree of condensation $C_{d}$ increases and correspondingly von Neumann entropy $S_{1}$ decreases as shown in Fig.~\ref{fig:l0_qent}, leading to increase in quantum mechanical phase coherence.
The explanation for such a behavior is that as the range $\sigma$ of the Gaussian potential is increased, the broadened interparticle pair-potentials begin to overlap leading to increase in many-body effect.
This increased many-body effect compared to the zero-range ($\delta$-function) potential leads to an enhanced phase coherence, reflected in Figs.~\ref{fig:l0_qent} and \ref{fig:l0_doc}.
Thus the finite-range (Gaussian) potential leads to Bose-Einstein condensate with relatively lower value of entropy and higher degree of condensation as compared to zero-range ($\delta$-function) potential \cite{lgd12}.
We further show, in Fig.~\ref{fig:qentdoc}, that how the von Neumann entropy $S_{1}$ gains a large value and the degree of condensation $C_{d}$ decreases with number of particles
{\it i.e.} for a given value of $\sigma$, the fragmentation is larger for larger number of bosons $N$ \cite{msp17,kmk17}.
A fragmented state is defined as a state for which there exits more than one macroscopically large eigenvalue of the SPRDM \cite{po56,nj82,mhu06}.

\subsection{Rotating System}
The results presented so far have been for the non-rotating $L_{z}=0$ states.
As the system is subjected to an external rotation, higher angular momentum states $\left(L_{z} > 0\right)$ which minimize the free energy in the co-rotating frame, become the ground state \cite{wg00}.
We now examine the ground state properties of the rotating system as a function of the interaction range $\sigma$ of the repulsive Gaussian potential in Eq.~(\ref{gip}), for a fixed number of bosons $N$ in different subspaces of quantized total angular momentum $L_{z}$.
For the sake of comparison, the results are also presented for the non-rotating state $L_{z}=0$ in the respective plots. 

\begin{figure}[!htb]
\subfigure[~$N=4$; $L_{z}=0,2,4,6,8$]{\includegraphics[width=0.49\linewidth]{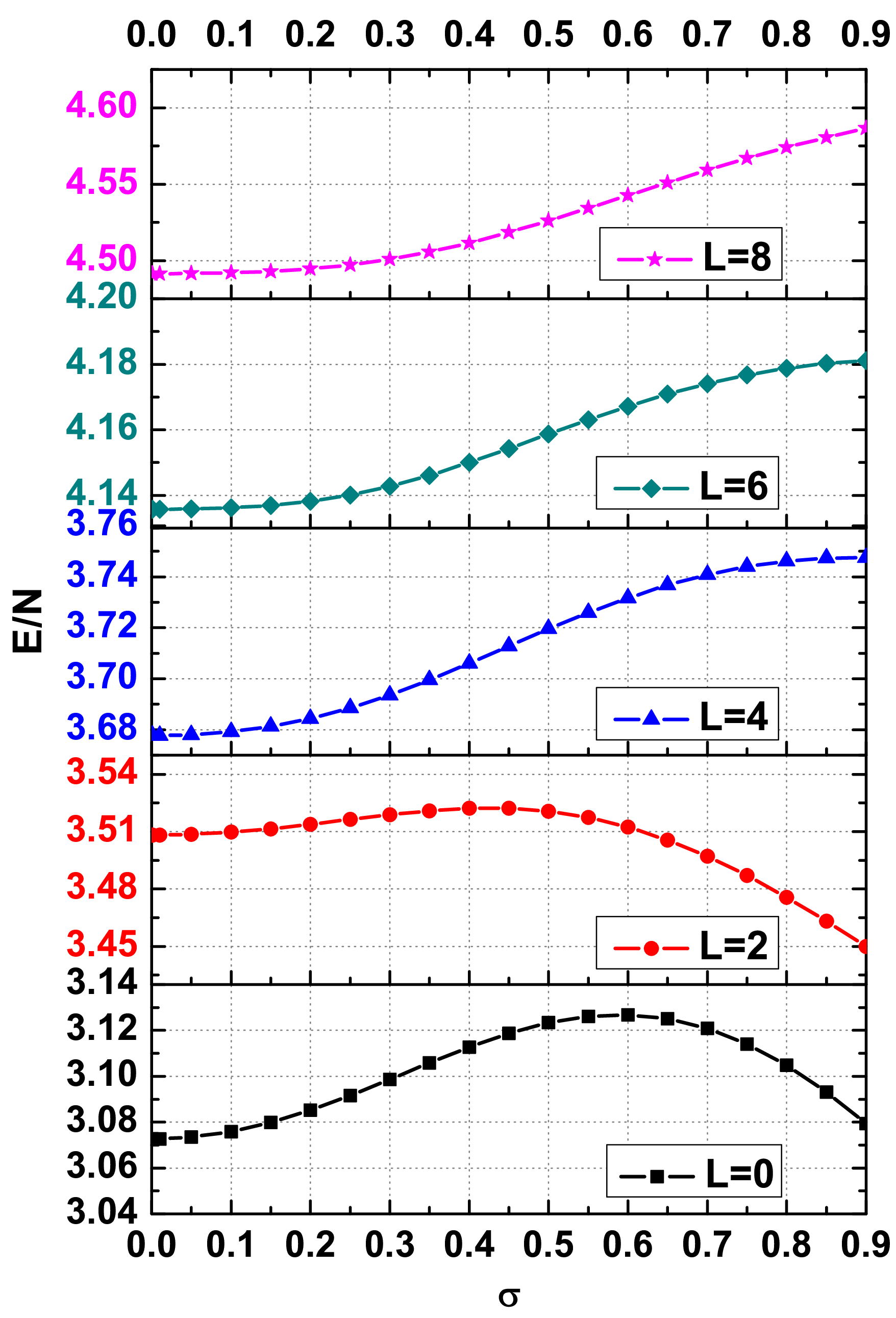}\label{fig:n4_ens}}
\hglue -1mm
\subfigure[~$N=8$; $L_{z}=0,4,8,12,16$]{\includegraphics[width=0.49\linewidth]{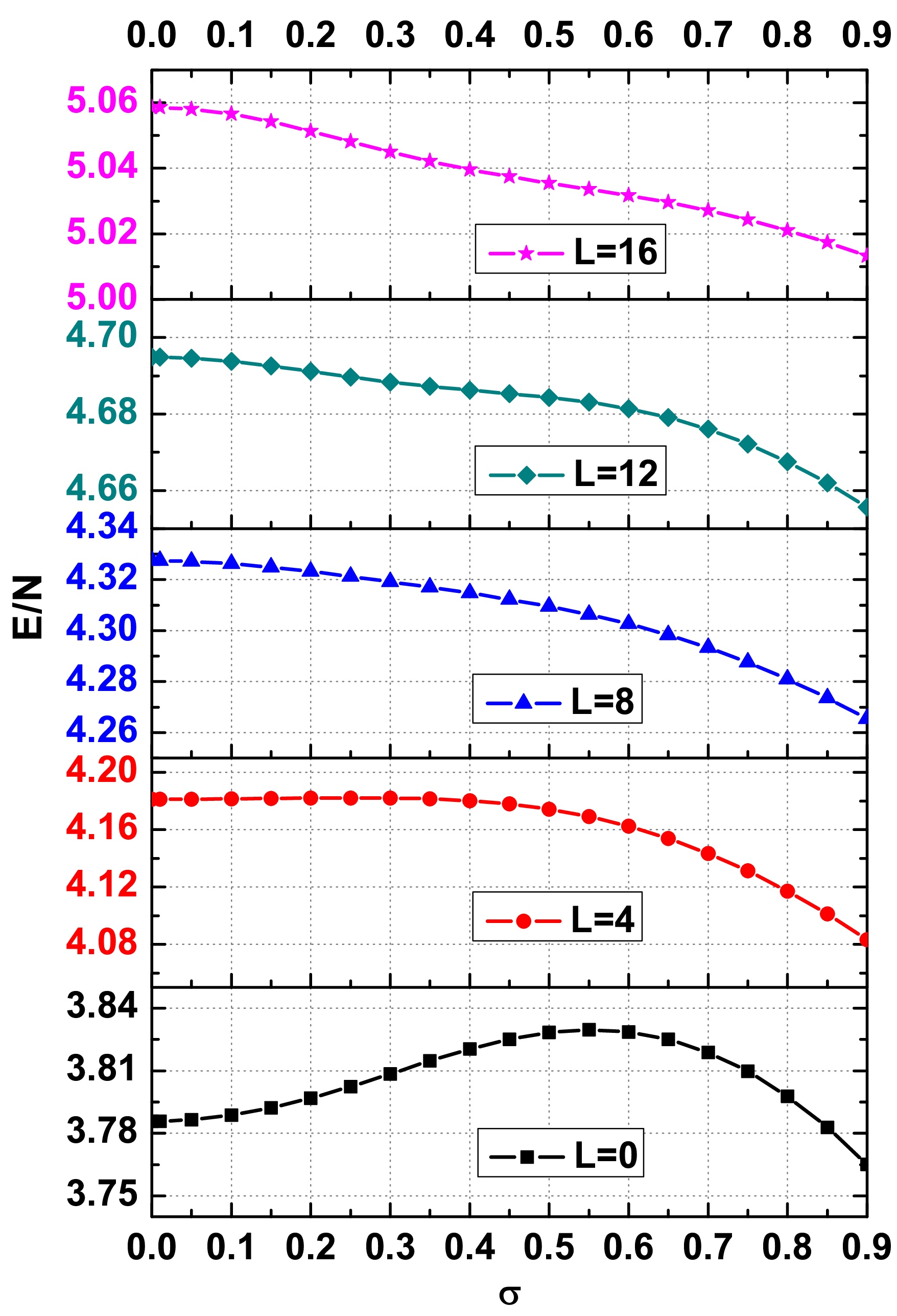}\label{fig:n8_ens}}
\caption{(Color online) The ground state energy per particle $E/N$ (in units of $\hbar \omega_{\perp}$) {\it versus} the interaction range $\sigma$ (in units of $a_{\perp}$), in different subspaces of $L_{z}$, with interaction parameter $\mbox{g}_{2}=9.151$ of the Gaussian potential~(\ref{gip}) for a rotating system of (a) $N=4$ and (b) $N=8$ bosons. The black line (squares) represents the ground state energy in $L_{z}=0$ subspace of the respective non-rotating system, included here for reference.}
\end{figure}

\begin{figure}[!htb]
\subfigure[~$S_{1}$ {\it versus} $\sigma$]{\includegraphics[width=0.9\linewidth]{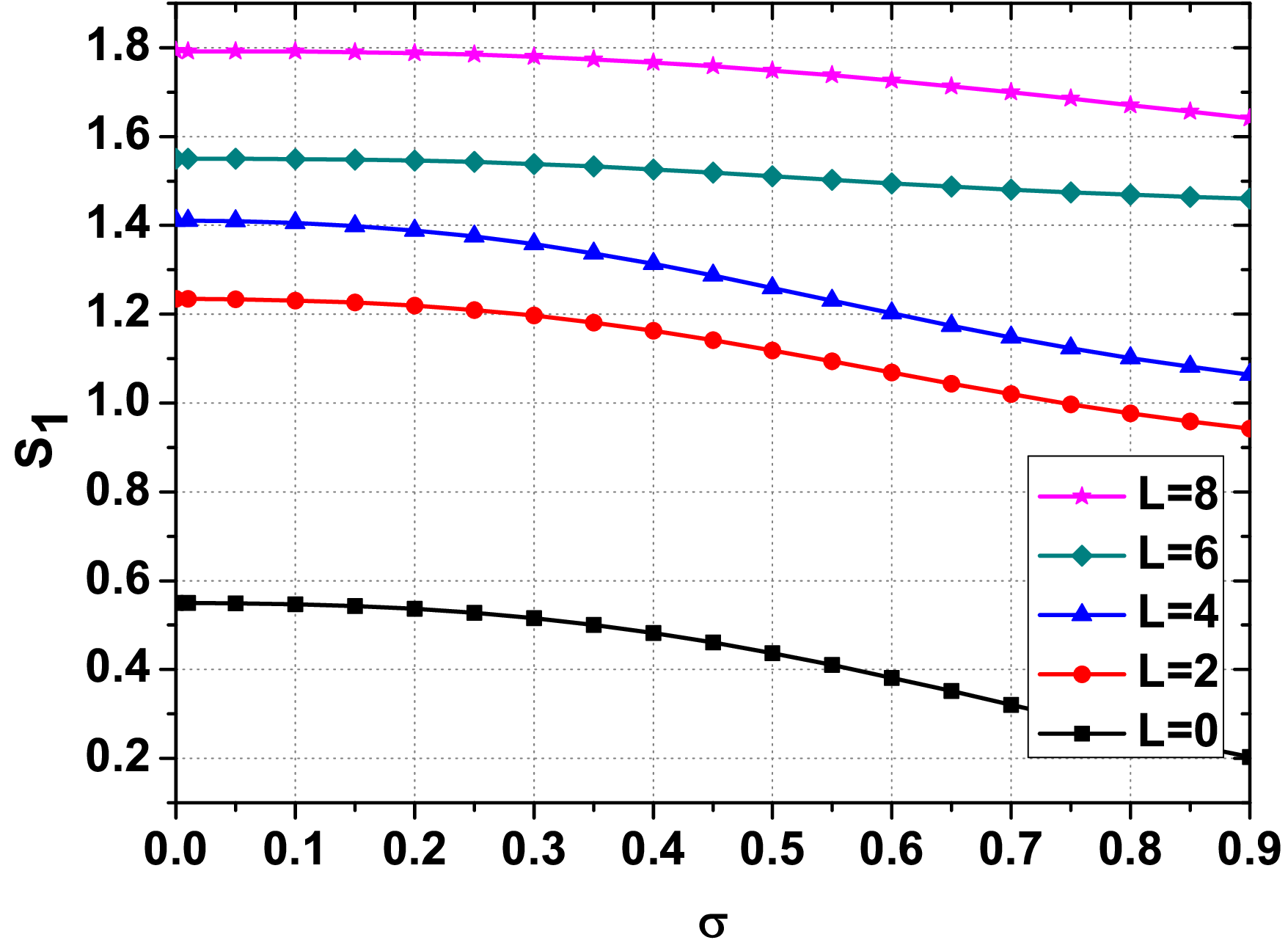}\label{fig:n4_qent}}
\subfigure[~$C_{d}$ {\it versus} $\sigma$]{\includegraphics[width=0.9\linewidth]{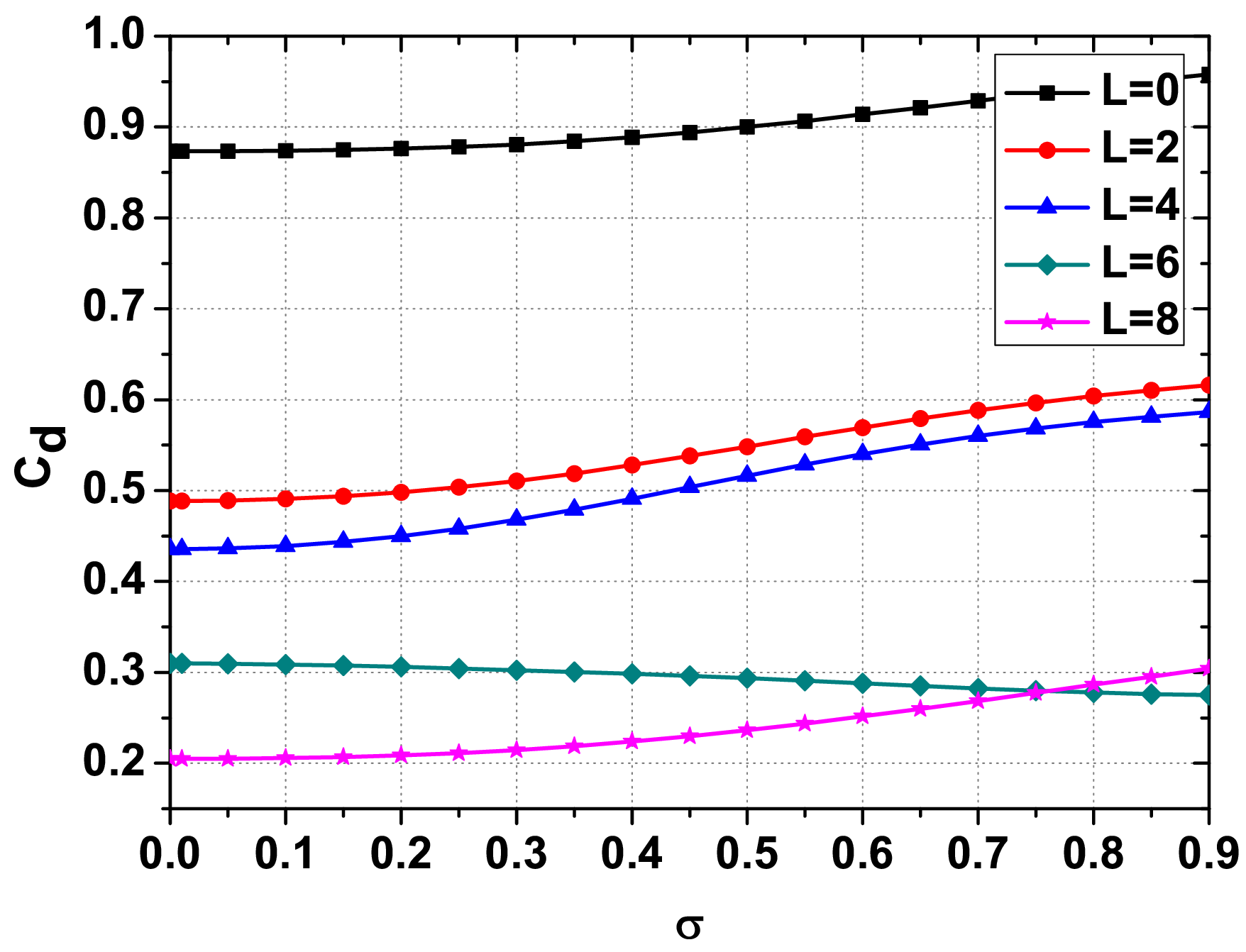}\label{fig:n4_doc}}
\caption{\label{fig:n4_qd}(Color online) For a rotating system of $N=4$ bosons in subspaces $L_{z}=0$~(black squares), $2$~(red circles), $4$~(blue triangles), $6$~(green rhombus) and $8$~(pink stars), the variation of (a) von Neumann entropy $S_{1}$ and (b) degree of condensation $C_{d}$ of the many-body quantum state, as a function of interaction range $\sigma$ (in units of $a_{\perp}$) of the Gaussian potential~(\ref{gip}) with interaction parameter $\mbox{g}_{2}=9.151$. The black line (squares) exhibit the results for the non-rotating system with $L_{z}=0$, included here for reference.}
\end{figure}

\paragraph*{Many-body state energy.}
For a system of $N=4$ and $N=8$ bosons, Figures~\ref{fig:n4_ens} and \ref{fig:n8_ens} respectively, present the variation of the lowest state energy per particle $(E/N)$ in the co-rotating frame with the interaction range $\sigma$, in different subspaces of total angular momentum $L_{z}$.
First, we consider a few-body system with $N=4$ and observe from Fig.~\ref{fig:n4_ens} that the $\sigma$-dependence of the lowest state energy per particle $E/N$ corresponding to the state $L_{z}=2$ is similar to the non-rotating ground state with $L_{z}=0$. 
As the angular momentum is increased beyond $L_{z}=2$, the energy per particle $E/N$ increases, in general, linearly with $\sigma$.
For small values of $\sigma\ (< 0.1)$, the value of $E/N$ does not change significantly with $\sigma$.
However, for large values of $\sigma\ (> 0.1)$, the effect of increasing rotation (increasing $L_{z}$) on $E/N$ dominates over the effect of $\sigma$.
When the particle number is increased to $N=8$, the ground state energy $E/N$ decreases with increasing $\sigma$ for a given angular momentum ($L_{z} \neq 0$) and becomes almost linear for higher angular momentum states, as seen in Fig.~\ref{fig:n8_ens}. 
The trend is thus opposite to the one observed for $N=4$ system and can be attributed to quantum (Bose-Einstein) statistics.
\paragraph*{Quantum correlation.}
For a rotating system with a fixed number of bosons $N$, we calculate von Neumann entanglement entropy $S_{1}$~(\ref{qent}) and degree of condensation $C_{d}$~(\ref{doc}), in different subspaces of total angular momentum $L_{z}$.
In Fig.~\ref{fig:n4_qd}, we plot $S_{1}$ and $C_{d}$ as a function of interaction range $\sigma$, for a few-boson system with $N=4$ in subspaces $L_{z}=0,\,2,\,4,\,6,\,8$. 
It is observed from Fig.~\ref{fig:n4_qent} that for a given $\sigma$, the von Neumann entropy $S_{1}$ increases with increase in $L_{z}$.
Further, in a given subspace of $L_{z}$, $S_{1}$ decreases for increasing $\sigma$ specifically at large values of $\sigma$.
The decreasing behavior of $S_{1}$ with increasing $\sigma$, is relatively more pronounced for $L_{z}=0,2,4,8$ corresponding to no vortex, single vortex and two vortex states respectively, and moderate for the fragmented state $L_{z}=6$, indicating thereby that the effect of rotation on $S_{1}$ is similar to the one observed with energy, as discussed above in Fig.~\ref{fig:n4_ens} for $N=4$ bosons. 
It is evident from Fig.~\ref{fig:n4_doc} that the degree of condensation $C_{d}$ increases gradually with increasing $\sigma$ for a given $L_{z}$, except for the fragmented state $L_{z}=6$, in which it is decreasing with increasing $\sigma$.
Moreover, the curves $C_{d}$ {\it versus} $\sigma$ for the subspaces $L_{z}=6$ and $L_{z}=8$ intersect at $\sigma=0.75$, implying the same value of degree of condensation.
We find that over the range $0 \le \sigma \leq 0.9$, the angular momentum state $L_{z}=6$ is a fragmented state with more than one macroscopically large eigenvalues $\lambda_{1} \sim \lambda_{2}$ of the SPRDM in Eq.~(\ref{spd}). 
For example, at $\sigma=0.75$ in $L_{z}=6$ subspace for $N=4$, the largest two eigenvalues of SPRDM are $\lambda_{1}=0.340$ and $\lambda_{2}=0.310$.
On the other hand, at $\sigma=0.75$ in $L_{z}=8$ subspace corresponding to second vortex state for $N=4$, the largest two eigenvalues of SPRDM are $\lambda_{1}=0.338$ and $\lambda_{2}=0.213$. 
In summary, the effect of interaction range $\sigma$ and rotation are two competing effects; however, the effect of $\sigma$ is overwhelmed by the effect of rotation since the energy scale of rotation {\it i.e.} $\hbar\Omega$ is large compared to the energy scale of interaction involving $\sigma$. Consequently, the effect of increasing $\sigma$ in enhancing the quantum phase coherence (observed most pronounced in the non-rotating state with $L_{z}=0$), is gradually destroyed by rotation with increasing $L_{z}$ leading to the formation of fragmented state.
\begin{figure}[!htb]
\subfigure[~$S_{1}$ {\it versus} $\sigma$]{\includegraphics[width=0.9\linewidth]{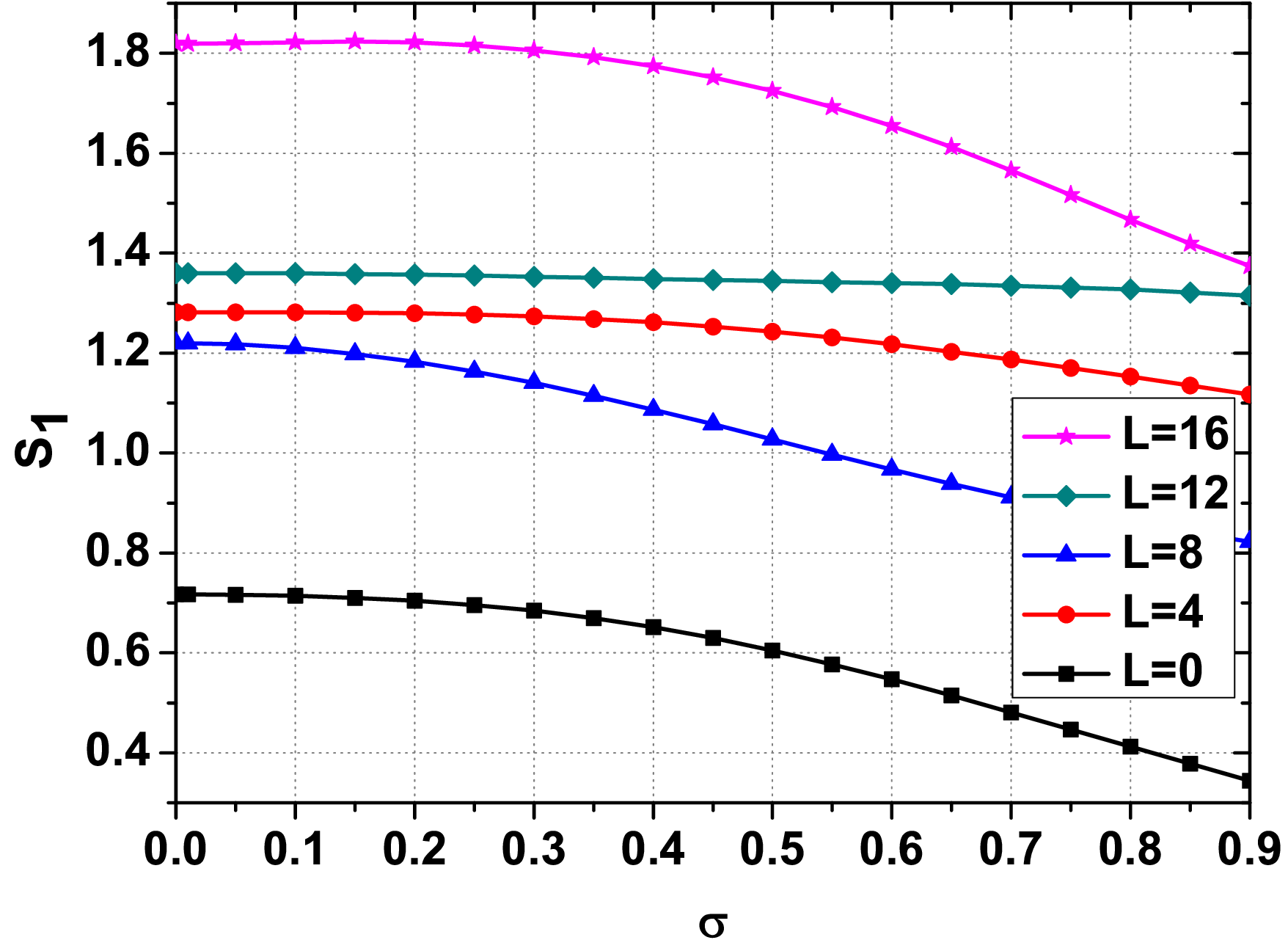}\label{fig:n8_qent}}
\subfigure[~$C_{d}$ {\it versus} $\sigma$]{\includegraphics[width=0.9\linewidth]{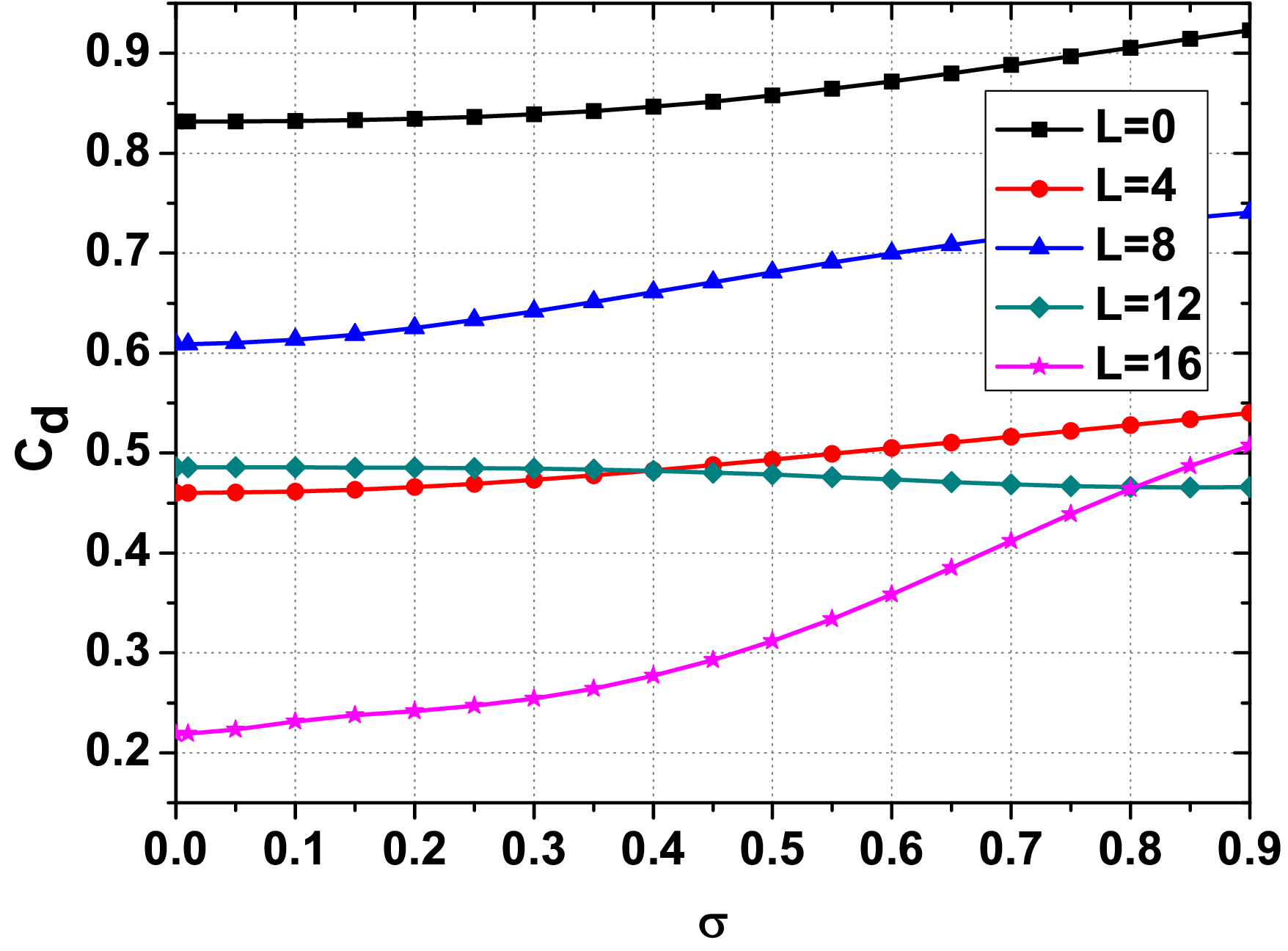}\label{fig:n8_doc}}
\caption{\label{fig:n8_qd}(Color online) The variation of (a) von Neumann entropy $S_{1}$ and (b) degree of condensation $C_{d}$ with interaction range $\sigma$ (in units of $a_{\perp}$) of the Gaussian potential in Eq.~(\ref{gip}) for a system of $N=8$ bosons in subspaces of $L_{z}=0$~(black squares), $4$~(red circles), $8$~(blue triangles), $12$~(green rhombus) and $16$~(pink stars); interaction parameter is chosen to be $\mbox{g}_{2}=9.151$. The black line (squares) exhibit the results for the non-rotating system with $L_{z}=0$, included here for reference.}
\end{figure}
\\
\indent 
We now consider a system of $N=8$ bosons which appears to be large enough to exhibit the effect of quantum statistics leading to nucleation of first and second vortex states \cite{br99,kmp00}.
We present in Fig.~\ref{fig:n8_qd} the variation of von Neumann entropy $S_{1}$ and degree of condensation $C_{d}$ with the interaction range $\sigma$, for $N=8$ bosons in different subspaces of $L_{z}$.
In Fig.~\ref{fig:n8_qent}, we observe that $S_{1}$ decreases with increasing value of $\sigma$ in each subspace of $L_{z}$.
This decreasing behavior of $S_{1}$ with $\sigma$ is found to be steep for $L_{z}=0,\ 8,\ 16$, moderate for $L_{z}=4$ and approximately constant for $L_{z}=12$.
It is to be noted from the figure that for a given value of $\sigma$, the von Neumann entropy $S_{1}$ increases with increasing angular momenta $L_{z}$, except for the first vortex state.
The value of $S_{1}$ is lower for the first vortex state with $L_{z}=N=8$ compared to the state with $L_{z}=4$. 
This is further supported by the fact that for a given $\sigma$, the degree of condensation $C_{d}$ decreases with increasing $L_{z}$, except for the first vortex state $L_{z}=N=8$ and the second vortex state $L_{z}=12$, as seen in Fig.~\ref{fig:n8_doc}.
Thus, for small values of $\sigma \le 0.4$, the $C_{d}$ {\it versus} $\sigma$ curve for $L_{z}=8$ and $L_{z}=12$  vortical states shift to higher values of $C_{d}$ compared to $L_{z}=4$ state.
However, for $L_{z}=12$ state, we notice that the interaction range $\sigma$ of the Gaussian potential has hardly any effect on $S_{1}$ and $C_{d}$.
Further, the $C_{d}$ versus $\sigma$ curve for the second vortical state $L_{z}=12$ intersects the $C_{d}$ curves for $L_{z}=4$ state at $\sigma=0.4$ and $L_{z}=16$ state at $\sigma=0.8$.
In the following subsection, we discuss the effect of $\sigma$ on the critical angular velocity of the first vortex state with $L_{z}=N$. 

\subsection{First Vortex State}
The non-rotating ground state of the system corresponds to total angular momentum $L_{z}=0$.
As the system is rotated, other (energetically favored) non-zero angular momentum states $\left(L_{z}\neq 0\right)$ successively become the ground state of the system \cite{wg00}.
These are obtained by minimizing the energy in the co-rotating frame \cite{ahs01}
\begin{equation}
E^{rot}\left(\Omega,L_{z},\mbox{g}_{2}\right)=E^{lab}\left(L_{z},\mbox{g}_{2}\right)-\Omega L_{z}
\end{equation}
where $\mbox{g}_{2}$ is the interaction parameter and $\Omega$ is the externally impressed angular velocity.
The energy of a vortex-free state $E^{rot}_{0}(\Omega)$ in the rotating frame is numerically equal to the energy $E^{lab}_{0}$ in the laboratory frame because the angular momentum of a vortex-free state is zero. 
A singly quantized vortex along the trap axis has total angular momentum $L_{z}=N$ \cite{mcw00,br99}; therefore, the corresponding energy of the system in the rotating frame is $E^{rot}_{1}(\Omega)= E^{lab}_{1}-\Omega N$.
The difference between these two energies is
\begin{equation}
\Delta E^{rot}(\Omega)=E^{rot}_{1}-E^{rot}_{0}=E^{lab}_{1}-\Omega N -E^{lab}_{0}. 
\end{equation}
Setting $\Delta E^{rot}(\Omega)\mid_{_{\Omega_{\bf c1}}}=0$ gives the critical angular velocity of the first vortex state with $L_{z}=N$,
\begin{equation}
\Omega_{\bf c1}=\frac{E^{lab}_{1}-E^{lab}_{0}}{N}
\end{equation}
expressed in terms of the energy $E^{lab}$ of the Bose-condensed state with and without a vortex, obtained in the laboratory frame.
\begin{figure}
{\includegraphics[width=0.98\linewidth]{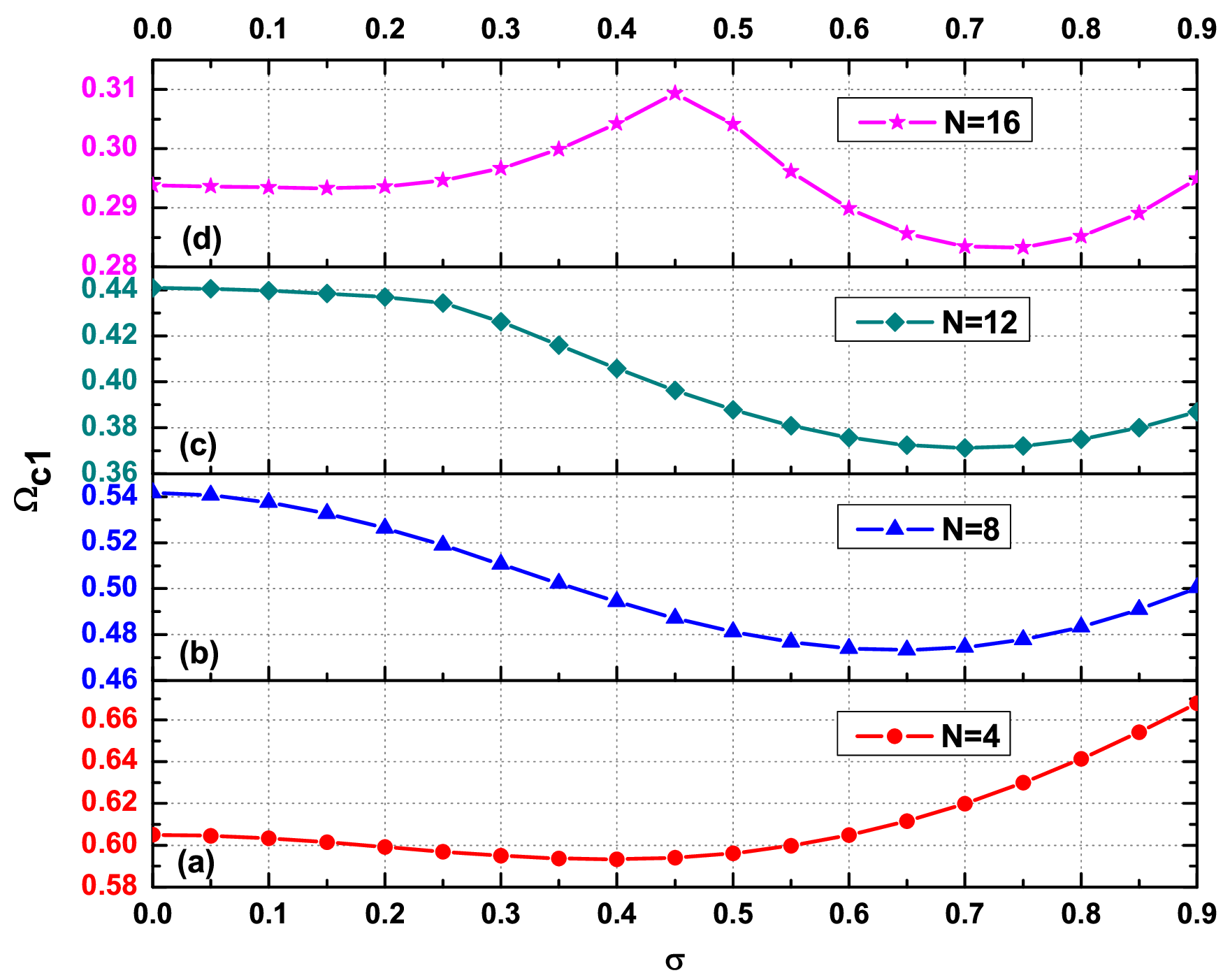}}
\caption{\label{fig:nlcr}(Color online) For $N$ trapped bosons, critical angular velocity $\Omega_{\bf c1}$ (in units of $\omega_{\perp}$) of the first centered vortex state ($L_{z}=N$) {\it versus} the interaction range $\sigma$ (in units of $a_{\perp}$) of the Gaussian potential~(\ref{gip}). With fixed value of interaction parameter $\mbox{g}_{2}=9.151$, the minimum value of $\Omega_{\bf c1}$ is observed (a) for $N=4$ at $\sigma=0.4$, (b) for $N=8$ at $\sigma=0.65$, (c) for $N=12$ at $\sigma=0.7$ and (d) for $N=16$ at $\sigma=0.75$ on the system size scale.}
\end{figure}
\\
\indent 
In Fig.~\ref{fig:nlcr}, we plot the critical angular velocity $\Omega_{\bf c1}$ of the first vortex state $\left(L_{z}=N\right)$ as a function of the interaction range $\sigma$ for a rotating system of $N=4,\,8,\,12$ and $16$ bosons. 
As is observed from the figure, for a given set of interaction parameters ($\mbox{g}_{2}$ and $\sigma$) of the repulsive Gaussian potential in Eq.~(\ref{gip}), the critical angular velocity $\Omega_{\bf c1}$ of the first vortex state ($L_{z}=N$) decreases with increase in number of bosons $N$ \cite{ahs01}.
It is further observed that as $\sigma$ is increased, the critical angular velocity $\Omega_{\bf c1}$, decreases over a wide range of $\sigma$, regardless of the number of bosons $N$. 
Thus, for a rotating system of $N$ bosons with a given interaction strength $\mbox{g}_{2}$, there is a value of $\sigma$ for which the critical angular velocity $\Omega_{\bf c1}$ takes a minimum value.
This implies that for an optimal value of $\sigma$ of the Gaussian potential, the nucleation of the first  vortex may begin at a lower value of rotational angular velocity $\Omega$ as compared to the $\delta$-function $(\sigma \rightarrow 0)$ potential.
\\
\indent
We notice from the figure that for small values of $\sigma~(<0.1)$, there is little change in the value of critical angular velocity $\Omega_{\bf c1}$ compared to its value for the $\delta$-function potential $(\sigma \rightarrow 0)$; as the value of $\sigma$ is increased further, the change in $\Omega_{\bf c1}$ becomes significant depending on the number of bosons $N$ considered.
For instance, as shown in Fig.~\ref{fig:nlcr}(a) for $N=4$ bosons, the critical angular velocity $\Omega_{\bf c1}$ decreases for intermediate values of $\sigma$ and then increases steeply for large values of $\sigma$. 
However, as seen in Fig.~\ref{fig:nlcr}(b) for $N=8$ (the least number of bosons in our present study where the effect of quantum statistics becomes significant), the decrease in value of $\Omega_{\bf c1}$ with $\sigma$ is steeper than for the $N=4$ boson system. 
Similarly, beyond $\sigma=0.25$, the decrease of $\Omega_{\bf c1}$ with $\sigma$ for $N=12$ is steeper than for $N=4$ as shown in Fig.~\ref{fig:nlcr}(c). 
However, for $N=16$ bosons shown in Fig.~\ref{fig:nlcr}(d), $\Omega_{\bf c1}$ increases initially with $\sigma$; reaches a maxima at $\sigma=0.45$ and then falls sharply with a minima at $\sigma=0.75$. 
The different shape of $\Omega_{\bf c1}$ {\it versus} $\sigma$ curve for $N=16$, compared to $N=12,\ 8,\ 4$, can be attributed to quantum statistics (and may be seen as a signature of an impending quantum phase transition).
It is also observed that the critical angular velocity $\Omega_{\bf c1}$ always increases for large values of $\sigma (<1)$, regardless of the number of bosons $N$ considered.
Moreover, with increase in number of bosons, the minimum value of $\Omega_{\bf c1}$ obtained with repulsive Gaussian potential shifts to larger values of $\sigma$. 
\section{Summary and Conclusion}
\label{conc}
In summary, we have presented an exact diagonalization study of the effect of repulsive Gaussian type interaction with finite-range $\sigma$ and large $s$-wave scattering length, on the rotating system of a finite number $\left(2 \le N \le 16\right)$ of harmonically confined bosons in different subspaces of quantized total angular momentum $L_{z}$.
The study considers variation with $\sigma$ of several  
physical quantities (that are directly or indirectly accessible experimentally) such as lowest eigenstate energy, critical angular velocity (for nucleation of vortices) and quantum correlation (measured in terms of von Neumann entanglement entropy and degree of condensation) of the trapped Bose-condensed gas.
The value of $\sigma$ is varied over from zero (corresponding to $\delta$-function interaction) to a value determined by the system size (measured in units of $a_{\perp}$).
We explore the role of interaction range on the ground state properties of small systems ($N=2$ to $8$) where an understanding of few-body effect may provide valuable insight into large systems ($N=8$ to $16$) which admit the possibility of many-body correlation.
\\
\indent
For the non-rotating ($L_{z}=0$) system of $N=2,4,8,12,16$ bosons, the variation of ground state energy with interaction range $\sigma$ is analyzed.  
It is found that for small values of $\sigma$, the ground state energy increases for few-boson ($2\le N\le 8$) system but decreases for many-boson ($N>8$) system.
On the other hand for relatively large values of $\sigma\ (<1)$, the ground state energy exhibits a monotonic decrease, regardless of the number of bosons $N$, with the system reducing to a droplet of bosons \cite{srh10} where quantum statistics ceases to play any significant role. 
Further, for a given $N$, with increasing $\sigma$, the value of von Neumann entropy decreases and the degree of condensation increases.
Thus, one achieves a more phase-rigid Bose-Einstein condensate with relatively lower value of entropy and higher degree of condensation using repulsive finite-range Gaussian potential compared to zero-range ($\delta$-function) potential.
\\
\indent
For a rotating system in the angular momentum regime $0 \le L_{z}\le 2N$, it is observed that the effect of rotation (with quantized $L_{z}$ for a given $N$) competes with the effect of interaction range $\sigma$ on the ground state properties of the Bose-condensate, though the former effect dominates over the latter effect.
In a given subspace of $L_{z}(>0)$, the value of lowest energy with $\sigma$ increases gradually for few boson system ($N=4$) but decreases for a system like $N=8$ where the quantum statistics begins to show up.
This observation supports the notion that when the particle number is increased to $N=8$, the $\sigma$-dependence of the energy follows a trend opposite to that of $N=4$ and can be attributed to quantum (Bose-Einstein) statistics.
For a given $L_{z}$, the quantum correlation is enhanced with increasing $\sigma$ (similar to the non-rotating  $L_{z}=0$ state) as the von Neumann entropy decreases and degree of condensation increases, specifically at large values of $\sigma$.
However, for a given $\sigma$, the von Neumann entropy increases and degree of condensation decreases with increase in $L_{z}$ {\it i.e.} rotation. 
\\
\indent
We further observe that for a given interaction range $\sigma$, the critical angular velocity $\Omega_{\bf c1}$ of the single vortex state $\left(L_{z}=N\right)$ decrease with increase in number of bosons $N$ \cite{ahs01}.
However, an increase in $\sigma$ leads, in general, to a smaller critical angular velocity $\Omega_{\bf c1}$ regardless, of the number of bosons $N$.
For a given number $N$ of bosons, there is a value of $\sigma$ for which the critical angular velocity $\Omega_{\bf c1}$ attains a minimum value indicating that there is an optimal value of $\sigma$ of the Gaussian potential for which the nucleation of the first vortex begins at a lower value of rotational angular velocity compared to the zero-range $\delta$-function potential.
With increase in number of bosons $N$, the minimum value of $\Omega_{\bf c1}$ obtained with Gaussian potential shifts to larger values of $\sigma$.
\\
\indent
Thus, we find that in a quasi-2D system, the effect of interaction range of the repulsive Gaussian potential on the ground state properties is negligible for small values of $\sigma (<0.1)$.
However, the effect becomes significant for values of $\sigma$ over the range $1>\sigma >0.1)$ on the system size scale $a_{\perp}$.
The results of the present study demonstrates the suitability of finite-range Gaussian interaction as a model for two-body interaction in a quantum many-body Bose system.
With the possibility of experimental realization of few-body quantum system and the experimental controllability of the interatomic interaction via Feshbach resonance \cite{ias98,cgj10}, the present theoretical work on atomic gases will be useful for the study of rotating quantum many-body systems with more realistic interaction potential in two dimensions.


\begin{thebibliography}{}
\bibitem{aem95} M. H. Anderson, J. R. Ensher, M. R. Matthews, C. E. Wieman and E. A. Cornell, {Science}, \textbf{269}, 198 (1995). % {Rb}
\bibitem{dma95} K. B. Davis, M.-O. Mewes, M. R. Andrews, N. J. van Druten, D. S. Durfee, D. M. Kurn and W. Ketterle,  {Phys. Rev. Lett.} \textbf{75}, 3969 (1995). % {Na}
\bibitem{bst95} C. C. Bradley, C. A. Sackett, J. J. Tollett and R. G. Hulet, {Phys. Rev. Lett.} \textbf{75}, 1687 (1995). %{Li}
\bibitem{isw99} See, Proceedings of the International School of Physics ``Enrico Fermi", Course CXL, 1999, edited by M. Inguscio, S. Stringari, and C. E. Wieman (IOS Press, Netherlands, 1999).
\bibitem{dgp99} F. Dalfovo, S. Giorgini, L. P. Pitaevskii, and S. Stringari, Rev. Mod. Phys. \textbf{71}, 463 (1999).
\bibitem{fs01} A. L. Fetter and A. A. Svidzinsky, {J. Phys.-Cond. Matt.} \textbf{13}, R135 (2001).
\bibitem{ajl01} A. J. Leggett, Rev. Mod. Phys. \textbf{73}, 307 (2001).
\bibitem{ps02} C. J. Pethick and H. Smith, \textit{Bose-Einstein Condensation in Dilute Gases}, (Cambridge University Press, Cambridge, England, 2002).
\bibitem{jfa03} J. F. Annett, \textit{Superconductivity, superfluids and condensates}, (Oxford University Press, Bristol, 2003). 
\bibitem{ias98} S. Inouye, M. R. Andrews, J. Stenger, H.-J. Miesner, D. M. Stamper-Kurn, and W. Ketterle, Nature \textbf{392}, 151 (1998).
\bibitem{cgj10} C. Chin, R. Grimm, P. Julienne, and E. Tiesinga, Rev. Mod. Phys. \textbf{82}, 1225 (2010).
\bibitem{bdz08} I. Bloch, J. Dalibard and W. Zwerger, {Rev. Mod. Phys.} \textbf{80}, 885 (2008).
\bibitem{coo08} N. R. Cooper, {Adv. Phys.} \textbf{57}, 539 (2008).
\bibitem{fet09} A. L. Fetter, {Rev. Mod. Phys.} \textbf{81}, 647 (2009).
\bibitem{mo06} O. Morsch and M. Oberthaler, Rev. Mod. Phys. \textbf{78}, 179 (2006).
\bibitem{bpt10} W. S. Bakr, A. Peng, M. E. Tai, R. Ma, J. Simon, J. I. Gillen, S. F{\"o}lling, L. Pollet, and M. Greiner, Science \textbf{329}, 547 (2010).
\bibitem{swe10} J. F. Sherson, C. Weitenberg, M. Endres, M. Cheneau, I. Bloch, and S. Khur, Nature (London) \textbf{467}, 68 (2010).
\bibitem{lsh03} R. Long, T. Steinmetz, P. Hommelhoff, W. H{\"a}nsel, T. W. H{\"a}nsch, and J. Reichel, Philos. Trans. R. Soc. London, Ser. A \textbf{361}, 1375 (2003).
\bibitem{fz07} J. Fort\'agh and C. Zimmermann, {Rev. Mod. Phys.} {\bf 79}, 235 (2007).
\bibitem{yl07} C. Yannouleas and U. Landman, {Rep. Prog. Phys.} {\bf 70}, 2067 (2007).
\bibitem{vie08} S. Viefers, {J. Phys.-Cond. Matt.} \textbf{20}, 123202 (2008).
\bibitem{srh10} H. Saarikoski, S. Reimann, A. Harju, and M. Manninen, {Rev. Mod. Phys.} \textbf{82}, 2785 (2010).
\bibitem{imp15} R. Islam, R. Ma, P. M. Preiss, M. E. Tai, A. Lukin, M. Rispoli, and M. Greiner, Nature (London) \textbf{528}, 77 (2015).
\bibitem{blu12} D. Blume, Rep. Prog. Phys. \textbf{75}, 046401 (2012).
\bibitem{msp17} Pere Mujal, Enric Sarl\'e and Artur Polls and Bruno Juli\'a-D\'{\i}az, Phys. Rev. A \textbf{96}, 043614 (2017).
\bibitem{kmk17} G.C. Katsimiga, S.I. Mistakidis, G.M. Koutentakis, P. G. Kevrekidis, P. Schmelcher, New J. Phys. \textbf{19}, 123012 (2017).
%
\bibitem{zgg18} Zhaoyang Zhang,  Ji Guo, Bingling Gu, Ling Hao, Gaoguo Yang, Kun Wang, Yanpeng Zhang, Photonics Research \textbf{6}, 713 (2018).
\bibitem{zzy15} Zhaoyang Zhang, Huaibin Zheng, Xin Yao, Yaling Tian, Junling Che, Xiuxiu Wang, Dayu Zhu, Yanpeng Zhang, Min Xiao, Scientific reports, \textbf{5}, 10462 (2015).
\bibitem{zwn11} Yanpeng Zhang, Zhiguo Wang, Zhiqiang Nie, Changbiao Li, Haixia Chen, Keqing Lu, Min Xiao, Phys. Rev. Lett. \textbf{106}, 093904 (2011).
\bibitem{ljz17} Changbiao Li, Zihai Jiang, Yiqi Zhang, Zhaoyang Zhang, Feng Wen, Haixia Chen, Yanpeng Zhang, Min Xiao, Phys. Rev. Applied \textbf{7}, 014023 (2017).
%
\bibitem{ber98} T. Busch, B.-G. Englert, K. Rzaewski, and M. Wilkens, Found. Phys. \textbf{28}, 549 (1998).
\bibitem{wg00} N. K. Wilkin and J. M. F. Gunn, {Phys. Rev. Lett.} {\bf 84}, 6 (2000).
\bibitem{cwg01} N. R. Cooper, N. K. Wilkin, and J. M. F. Gunn, Phys. Rev. Lett. \textbf{87}, 120405 (2001).
\bibitem{lhd10} X.-J. Liu, H. Hu, and P. D. Drummond, Phys. Rev. B \textbf{82}, 054524 (2010).
\bibitem{eg99} B. D. Esry and C. H. Greene, {Phys. Rev. A} \textbf{60}, 1451 (1999).
\bibitem{far10} A. Farrell and B. P. van Zyl, J. Phys. A \textbf{43}, 015302 (2010).
\bibitem{bp99} G. F. Bertsch and T. Papenbrock, {Phys. Rev. Lett.} \textbf{83}, 5412 (1999).
\bibitem{dka13} R. A. Doganov, S. Klaiman, O. E. Alon, A. I. Streltsov and L. S. Cederbaum, {Phys. Rev. A}, \textbf{87}, 033631 (2013).
\bibitem{cfa09} J. Christensson, C. Forss\'{e}n, S. \r{A}berg and S. M. Reimann, {Phys. Rev. A} \textbf{79}, 012707 (2009).
\bibitem{kls14} S. Klaiman, A. U. J. Lode, A. I. Streltsov, L. S. Cederbaum, and O. E. Alon, Phys. Rev. A \textbf{90}, 043620 (2014).
\bibitem{flc15} U. R. Fischer, A. U. J. Lode, and B. Chatterjee, Phys. Rev. A \textbf{91}, 063621 (2015).
\bibitem{iasl15} Mohd. Imran and M. A. H. Ahsan, {Adv. Sci. Lett.} \textbf{21}, 2764 (2015).
\bibitem{bkc15} R. Beinke, S. Klaiman, L. S. Cederbaum, A. I. Streltsov and O. E. Alon, Phys. Rev. A \textbf{92}, 043627 (2015).
\bibitem{sk16} K. Sakmann, and M. Kasevich, Nat. Phys. \textbf{12}, 451 (2016).
\bibitem{iactp16} Mohd. Imran and M. A. H. Ahsan, {Commun. Theor. Phys.} \textbf{65}, 473 (2016).
\bibitem{wtc17} S. E. Weiner, M. C. Tsatsos, L. S. Cederbaum, and A. U. J. Lode, Sci. Rep. \textbf{7}, 40122 (2017).
\bibitem{iajpb17} Mohd. Imran and M. A. H. Ahsan, {J. Phys. B: At. Mol. Opt. Phys.} \textbf{50}, 045301 (2017).
\bibitem{bca18} R. Beinke, L. S. Cederbaum, and O. E. Alon, Phys. Rev. A \textbf{98}, 053634 (2018).
\bibitem{jcb18} Peter Jeszenszki, Alexander Yu. Cherny and Joachim Brand, Phys. Rev. A \textbf{97}, 042708 (2018).
%
\bibitem{hua87} K. Huang, {\it Statistical Mechanics} (John Wiley \& Sons, New york, 1987). 
\bibitem{motiv} {The exact diagonalization of the Hamiltonian matrix for finite $N$-body problem yields an exact variational solution within the basis chosen. For reasons of computational feasibility, it becomes necessary to truncate the set of single-particle basis functions used in the construction of many-body variational wavefunction. The $\delta$-function potential $\mbox{g}_{2} \delta \left({\bf r} - {\bf r^{\prime}}\right) = \mbox{g}_{2} \sum_{\bf n} {u}_{\bf n}^{\ast}\left({\bf r}\right) {u}_{\bf n} \left({\bf r^{\prime}}\right)$, where summation $\bf{n}$ runs over all basis states, is expandable in an infinite set of single-particle basis and hence computationally not suitable for exact diagonalization \cite{hua87}.}
\bibitem {ahs01} M. A. H. Ahsan and N. Kumar, {Phys. Rev. A} \textbf{64}, 013608 (2001).
\bibitem{lhc01_pra} X.-J. Liu, H. Hu, L. Chang and S.-Q. Li, {Phys. Rev. A} \textbf{64}, 035601 (2001). %CPD
\bibitem{rot} {With rotational angular velocity much less than the trapping frequency, the degeneracy of the Landau levels is lifted even without interparticle interaction.}
\bibitem{int} {The particle-particle interaction causes different single-particle angular momentum states to scatter into each other.}
\bibitem{dav75} E. R. Davidson, {J. Comput. Phys.}, \textbf{17}, 87 (1975).
\bibitem{gok88} E. K. U. Gross, L. N. Oliveira and W. Kohn, {Phys. Rev. A} {\bf 37}, 2805 (1988).
\bibitem{dbo07} D. Dagnino, N. Barber\`{a}n, K. Osterloh, A. Riera, and M. Lewenstein, {Phys. Rev. A} \textbf{76}, 013625 (2007).
\bibitem{py01} R. Pa\ifmmode \check{s}\else \v{s}\fi{}kauskas, and L. You, {Phys. Rev. A} \textbf{64}, 042310 (2001).
\bibitem{lgc09} Z. Liu, H. Guo, S. Chen, and H. Fan, {Phys. Rev. A} {\bf 80}, 063606 (2009).
\bibitem{lf10} Z. Liu and H. Fan, {Phys. Rev. A} {\bf 81}, 062302 (2010).
%
\bibitem{lgd12} J. Y. Lee, X. W. Guan, A. del Campo, and M. T. Batchelor, Phys. Rev. A \textbf{85}, 013629 (2012).
\bibitem{po56} O. Penrose and L. Onsager, Phys. Rev. \textbf{104}, 576 (1956).
\bibitem{nj82} P. Nozieres and D. Saint James, J. Phys. (Paris) \textbf{43}, 1133 (1982).
\bibitem{mhu06} E. J. Mueller, T.-L. Ho, M. Ueda, and G. Baym, Phys. Rev. A \textbf{74}, 033612 (2006).
%
\bibitem{br99} D. A. Butts and D. S. Rokhsar, Nature (London) \textbf{397}, 327 (1999).
\bibitem{kmp00} G. M. Kavoulakis, B. Mottelson, and C. Pethick, Phys. Rev. A \textbf{62}, 063605 (2000).
\bibitem{mcw00} K. W. Madison, F. Chevy, W. Wohlleben, and J. Dalibard, Phys. Rev. Lett. \textbf{84}, 806 (2000).
\end{thebibliography}
\end{document}